\documentclass{aa} 
\usepackage{graphicx}
\usepackage{txfonts}
\begin{document}
\newcommand{\magcir}{\ \raise -2.truept\hbox{\rlap{\hbox{$\sim$}}\raise5.truept
 	\hbox{$>$}\ }}	
\newcommand{\mincir}{\ \raise -2.truept\hbox{\rlap{\hbox{$\sim$}}\raise5.truept
 \hbox{$<$}\ }} 

  \title{The Great Observatories Origins Deep Survey}
  \subtitle{VLT/VIMOS Spectroscopy in the GOODS-South Field}  

  \author{P. Popesso\inst{1}
     \and
      M. Dickinson\inst{4}
     \and 
      M. Nonino\inst{3}
      \and
      E. Vanzella\inst{2,3}
      \and
      E. Daddi\inst{8}
     \and 
      R.A.E. Fosbury\inst{5}
     \and 
      H. Kuntschner\inst{5}
      \and
      \\
      V. Mainieri\inst{7}
     \and
      S. Cristiani\inst{3}
     \and
      C. Cesarsky\inst{7}
     \and
      M. Giavalisco\inst{6}
     \and 
      A. Renzini\inst{2}
     \and
     \\
     the GOODS Team 
     }


  \institute{Max-Planck-Institut fur extraterrestrische Physik,, Giessenbachstrasse 2, 85748 Garching, Germany
    \and
    Dipartimento di Astronomia dell'Universit\`a di Padova,
    Vicolo dell'Osservatorio 2,
    I-35122 Padova, Italy.  
     \and
       INAF - Osservatorio Astronomico di Trieste, Via G.B. Tiepolo 11,
       40131 Trieste, Italy.
     \and 
       National Optical Astronomy Obs., P.O. Box 26732, Tucson, AZ 85726.
     \and 
       ST-ECF, Karl-Schwarzschild Str. 2, 85748 Garching, Germany.
     \and
       Space Telescope Science Institute, 3700 San Martin Drive,
       Baltimore, MD 21218. 
     \and
       European Southern Observatory, Karl-Schwarzschild-Strasse 2,
Garching, D-85748, Germany.
     \and 
     Universite' Paris-Sud 11, Rue Georges Clemenceau 15, Orsay, F-91405, France
     \and
     Jet Propulsion Laboratory, California Institute of Technology,
     MS 169-506, 4800 Oak Grove Drive, Pasadena, CA 91109
     \thanks{Based on observations made at the European Southern
Observatory, Paranal, Chile (ESO programme 170.A-0788 {\it The Great
Observatories Origins Deep Survey: ESO Public Observations of the
SIRTF Legacy/HST Treasury/Chandra Deep Field South.})}}



\abstract {} {We present the first results of the VIsible Multiobject
Spectrograph (VIMOS) ESO/GOODS program of spectroscopy of faint
galaxies in the Chandra Deep Field South (CDF-S). The program
complements the FORS2 ESO/GOODS campaign.} {3312 spectra have been
obtained in service mode with VIMOS at the ESO/VLT UT3. The VIMOS
LR-Blue and MR grisms have been used to cover different redshift
ranges. Galaxies at $1.8 < z < 3.5$ have been observed in the GOODS
VIMOS-LR-Blue campaign.  Galaxies at $z < 1$ and Lyman Break Galaxies
at $z > 3.5$ have been observed in the VIMOS MR survey.}  {Here we
report results for the first 6 masks (out of 10 total) that have been
analyzed from each of the LR-Blue and MR grisms.  Spectra of 2344
spectra have been extracted from these 6 LR-Blue masks and 968 from 6
MR masks. 33\% of the LR-Blue and 18\% of the MR spectra are
serendipitous observations. We obtained 1481 redshifts in the LR-Blue
campaign and 656 in the MR campaign for a total success rate of 63\%
and 68\%, respectively, which increase to 70\% and 75\% when only the
primary targets are considered. By complementing our VIMOS
spectroscopic catalog with all existing spectroscopic redshifts
publicly available in the CDF-S, we created a redshift master
catalog. By comparing this redshift compilation with different
photometric redshift catalogs we estimate the completeness level of
the CDF-S spectroscopic coverage in several redshift bins.} {The
completeness level is very high, $> 60\%$, at $z < 3.5$, and it is
very uncertain at higher redshift.  The master catalog has been used
also to estimate completeness and contamination levels of different
galaxy photometric selection techniques, such as the BzK, the so
called 'sub'-U-dropout and the drop-out methods and to identify large
scale structures in the field.}

 
\authorrunning{P. Popesso et al.}
\maketitle
%

\section{Introduction}
The Great Observatories Origins Deep Survey (GOODS) is a public,
multi-facility project that aims at answering some of the most profound
questions in cosmology: how did galaxies form and assemble their
stellar mass? When was the morphological differentiation of galaxies
established and how did the Hubble Sequence form? How did AGN form and
evolve, and what role do they play in galaxy evolution? How much do
galaxies and AGN contribute to the extragalactic background light? A
project of this scope requires large and coordinated efforts from many
facilities, pushed to their limits, to collect a database of
sufficient quality and size for the task at hand. It also requires
that the data be readily available to the worldwide community for
independent analysis, verification, and follow-up.

The program targets
two carefully selected fields, the Hubble Deep Field North (HDF-N) and
the Chandra Deep Field South (CDF-S), with three NASA Great
Observatories (HST, Spitzer and Chandra), ESA's XMM-Newton, and a wide
variety of ground-based facilities. The area common to
all the observing programs is 320 arcmin$^2$, equally divided between
the North and South fields. For an overview of GOODS, see \cite{dick03}, 
\cite{renz02} and \cite{giava04a}. 

Spectroscopy is essential to reach the scientific goals of GOODS.
Reliable redshifts provide the time coordinate needed to delineate the
evolution of galaxy masses, morphologies, clustering, and star
formation. They calibrate the photometric redshifts that can be
derived from the imaging data at 0.36-8$\mu$m.  Spectroscopy will
measure physical diagnostics for galaxies in the GOODS field (e.g.,
emission line strengths and ratios to trace star formation, AGN
activity, ionization, and chemical abundance; absorption lines and
break amplitudes that are related to the stellar population
ages). Precise redshifts are also indispensable to properly plan for
future follow-up at higher dispersion, e.g., to study galaxy
kinematics or detailed spectral-line properties.

The ESO/GOODS spectroscopic program is designed to observe all
galaxies in the CDF-S field for which VLT optical spectroscopy is
likely to yield useful data. The program is organized in two
campaigns, carried out at VLT/FORS2 at UT1 and VLT/VIMOS at UT3. The
program makes full use of the VLT instrument capabilities, matching
targets to instrument and disperser combinations in order to maximize
the effectiveness of the observations. 

The FORS2 campaign is now completed (Vanzella et al. 2005, 2006,
2008). 1715 spectra of 1225 individual targets have been observed and
887 redshifts have determined as a result. Galaxies have
been selected adopting three different color criteria and using the
photometric redshifts.  The resulting redshift distribution typically
spans two redshift domains: from $z=0.5$ to 2 and $z=3$ to 6.5. The
reduced spectra and the derived redshifts are released to the
community through the ESO web page
http://www.eso.org/science/goods/.The typical redshift uncertainty is
estimated to be $\sigma_z \sim 0.001$.

We are currently conducting the VIMOS ESO/GOODS spectroscopic survey
to complement the FORS2 survey in terms of completeness and sky
coverage. The cumulative source counts on the CDF-S field taken from
the deep public FORS1 data (Szokoly et al. 2004), show that down to
$\rm{V_{AB}}=25$ mag there are ~6000 objects over the 160
$\rm{arcmin}^2$ of the GOODS field. Only the high multiplexing
capabilities of VIMOS at VLT can ensure to reach the required
completeness in a reasonable amount of time. On average $\sim 330$
objects at a time have been observed with the low resolution ($R\sim
250$) grism and $\sim 140$ with the medium ($R\sim 1000)$ grism.

A Medium resolution campaign in the red (VIMOS MR orange grism) is
designed to reach the required completeness in the $0.5 < z < 1.3$ and
$z > 3.5$ redshift ranges. A low resolution spectroscopic campaign is
conducted with the VIMOS LR-Blue grism to cover the $1.8 < z < 3.5$
redshift range not covered by the FORS2 spectroscopy. The aim is to
reach mag~$\sim24-25$ with adequate S/N, with this limiting magnitude
being in the B band for objects observed with the VIMOS LR-Blue grism,
in the R band for those observed in the VIMOS MR-Orange grism, and in
the z band for the objects observed with FORS2.

In this paper we report on the first 60\% of the VIMOS spectroscopic
follow-up campaign in the Chandra Deep Field South (CDF-S), carried
out with the VIMOS instrument at the ESO VLT from P74 to P78. 10 masks
have been observed in the LR-Blue grism and 10 with the MR Orange
grism. Here we report results for the first 6 masks that have been
analyzed from each of the LR-Blue and MR grisms.

The paper is organized as follows: in Sect. 2 we describe the survey
strategy and in Sect. 3 the observations and the data reduction. The
details of the redshift determination is presented in Sect. 4. In
Sect. 5 we discuss the data and in Sect. 6 the relaibility of the
photometric techniques used to identify the high redshift targets. In
Sec. 7 we present our the conclusions. Throughout this paper the
magnitudes are given in the AB system (AB~$\equiv 31.4 -
2.5\log\langle f_\nu / \mathrm{nJy} \rangle$), and the ACS F435W,
F606W, F775W, and F850LP filters are denoted hereafter as $B_{435}$,
$V_{606}$, $i_{775}$ and $z_{850}$, respectively. We assume a
cosmology with $\Omega_{\rm tot}, \Omega_M, \Omega_\Lambda = 1.0, 0.3,
0.7$ and $H_0 = 70$~km~s$^{-1}$~Mpc$^{-1}$.


\section{The survey strategy}

\subsection{The VIMOS instrument}
The VIsible MultiObject Spectrograph (VIMOS) is installed on the
ESO/VLT, at the Nasmyth focus of the VLT/UT3 'Melipal' (Le Fevre et
al. 2003). VIMOS is a 4-channel imaging spectrograph, each channel (a
'quadrant') covering $\sim 7 \times 8$ $\rm{arcmin}^2$ for a total
field of view (a 'pointing') of $\sim 218$ $\rm{arcmin}^2$. Each
channel is a complete spectrograph with the possibility to insert slit
masks $\sim 30 \times 30$ $\rm{cm}^2$ each at the entrance focal
plane, broad band filters or grisms to produce spectra on $2048 \times
4096$ $\rm{pixels}^2$ EEV CCD.

The pixel scale is 0.205 arcsec/pixel, providing excellent sampling of
the Paranal mean image quality and Nyquist sampling for a slit of 0.5
arcsec in width. The spectra resolution ranges from $\sim 200$ to
$\sim 5000$. Because the instrument field at the Nasmyth focus of the
VLT is large ($\sim 1$ m), there is no atmospheric dispersion
compensator. This requires the observer to limit observations to
airmasses below 1.1.

In the MOS mode of observations, short 'pre-images' are taken ahead of
the observing run. These are cross-correlated with the user-catalog to
match the instrument coordinate system to the atmospheric reference of
the user catalog. User masks are, then, prepared using the VMMPS tool,
provided by ESO, with an automated optimization of slit number and
position (see Bottini et al. 2005).

\begin{figure}
\begin{center}
\begin{minipage}{0.49\textwidth}
\resizebox{\hsize}{!}{\includegraphics{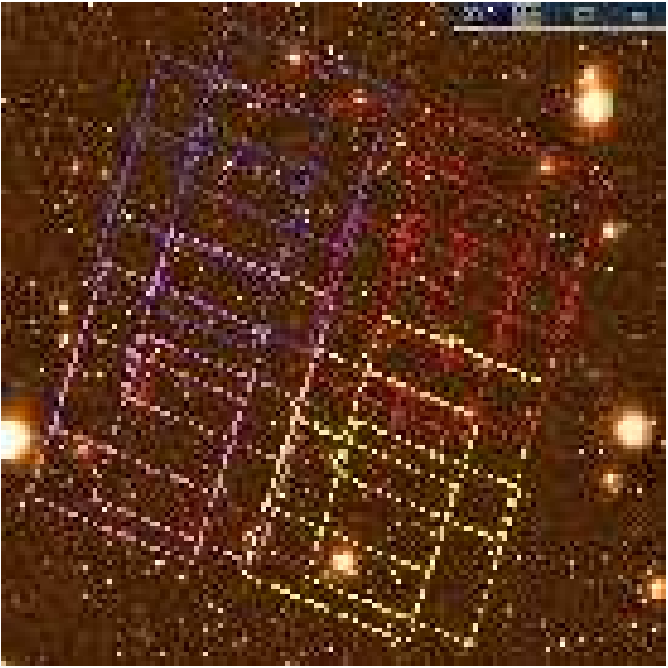}}
\end{minipage}
\end{center}
\caption{VIMOS field coverage of the GOODS area in the CDF-S.}
\label{cov}
\end{figure}

\subsection{The Field coverage: VIMOS pointing layout}
The VIMOS geometry (16' x 18', with a cross gap of 2' between the
quadrants) is such that only 50\% of it can overlap with the 10' x 16'
GOODS-CDFS field.  At least 3 VIMOS pointings are required to cover
the whole GOODS area (see Fig. \ref{cov}). The VIMOS multiplex allows
measurement of an average of $\sim 360$ per pointing in the case of
the Low Resolution (LR) grism and $\sim$ 150 in the case of the Medium
Resolution (MR) grism. 10 Low resolution and 10 Medium Resolution
masks (on average 3 LR and 3 MR masks per pointing) are necessary to
ensure $\sim 96$\% completeness in the spectroscopic coverage of $\sim
6000$ targets. With an average integration time of 4h per mask and
30\% overheads, this makes a total of 125h.

\subsection{Target Selection}

Several categories of object selection criteria have been used to
ensure a sufficiently high density of target candidates on the sky to
efficiently fill out multi-slit masks. The target selection is based
on the multiwavelength photometry available for the GOODS-S field.
Different criteria have been used for the low resolution 'LR-Blue'
grism and the medium resolution 'MR Orange' grism. In both part of the
GOODS VIMOS survey the selection criteria and the target catalog have
been tailored in time to optimize the survey success rate. The target
catalog have been changed according to the incoming new GOODS dataset:
partial results of the spectroscopic surveys, updated GOODS
photometric catalogs, Spitzer IRAC and MIPS data. Moreover, the VIMOS
quadrants cover an area larger than the GOODS field. Thus, different
datasets have been used to create the target catalogs of the GOODS
field and the remaining area. We summarize here the main criteria of
selection used during the GOODS VIMOS survey in ESO observing periods
P74-P78.

The VIMOS LR-Blue grism covers the wavelength range $3500-6900$
$\AA$. Hence it is suitable for the observation of the ultraviolet
absorption and emission features of objects in the redshift range $1.8
\le z \le 3.8$. The target selection of the low resolution campaign is
based on the following criteria:

\begin{itemize}

\item U-dropouts: CTIO and WFI photometry based U-drops (see Section 6
and Lee et al. 2006 for the detailed description of the selection
criteria)

\item BzK color-selected objects with IRAC color preselection (when available) to
favor $z\sim 2$ galaxies (Daddi et al. 2004)

\item ``so-called'' sub-U-dropouts: color-selected objects with $UBR$
  colors falling below of those of normal $z \sim 3$ Lyman break color
  selection criteria, similar to 'BX' selection criterion of Adelberger
  et al. (2004, see Section 6 for the detailed description of the
  selection criteria). Sub-U-dropouts selection criteria are based on
  CTIO U-band and WFI photometry.

\item X-ray sources from the CDF-S and E-CDF-S X-ray catalogs
(Giacconi et al. 2002, Lehmer et al. 2005)

\end{itemize}

No low redshift galaxies have been targeted for the LR-Blue masks.  A
magnitude cut at $B < 24.5$ mag is applied to all target catalogs listed
above.

The wavelength range of the VIMOS MR Orange grism is $4000-10000$
$\AA$, similarly to FORS2. However, the fringing at red wavelength
($\lambda \ge 7000$ $\AA$) is somewhat stronger than in FORS2. Hence,
the optical spectral emission and absorption features for galaxies at
$z > 1$, and the ultraviolet spectral features of Lyman break galaxies
(LBG) at $z \gtrsim 4.8$,which would appear at very red optical
wavelengths, are not well detectable. Therefore, the target selection
is limited to bright galaxies  and LBGs in the redshift range $2.8
< z < 4.8$. The selection is based, as for the LR-Blue campaign, on
the available photometry according to the following criteria:

\begin{enumerate}

\item galaxies at $R <24.5$ with the exclusion of VIMOS LR-Blue
  targets and objects already observed.

\item MIPS selected targets $R < 24.5$ mag with the exclusion of VIMOS
  LR-Blue targets and objects already observed.

\item faint Lyman break galaxies at $i_{775} < 25$, selected as
$B_{435}$, $V_{606}$ dropouts.

\end{enumerate}

When designing the masks, we avoid as much as possible to observe
targets that had already been observed in other redshift surveys of
this field, namely, the K20 survey of \cite{cimatti02}, the survey of
X-ray sources by Szokoly et al. (2004), the VIMOS VLT Deep Survey (Le
Fevre et al. 2005) and the ESO/GOODS FORS2 survey (Vanzella et
al. 2005, 2006).

\section{Observations and Data Reduction}
The VLT/VIMOS spectroscopic observations were carried out 
in service mode during ESO observing periods P74-P78. 

\subsection{Preparation of VIMOS observations} For each pointing a
short V-band image is taken with VIMOS ahead of the spectroscopic
observations. We used the preimaging togheter with the
GOODS WFI R band image to derive the transformation matrix from the
($\alpha$, $\delta$) sky reference frame of the target catalogs to the
($X_{CCD}$,$Y_{CCD}$) VIMOS instrumental coordinate system. The
cross-correlation is performed with the routines \emph{geomap} and
\emph{geoxytran} in the IRAF environment, which turn out to reach a
spatial accuracy, $\sim 0.05$ arcsec, ten times higher than the
accuracy of the matching procedure implemented in the VIMOS mask
preparation software (VMMPS, Bottini et al. 2005). This is due to the
choice of a higher order polynomial of the fitting procedure, which is
not allowed in VMMPS. Once the target catalog is expressed in the
($X_{CCD}$,$Y_{CCD}$) VIMOS instrumental coordinate system, the next
steps are conducted with VMMPS tool, to design the slit mask layout.
After placing two reference apertures on bright stars for each
pointing quadrant, slits are assigned to a maximum number of sources
in the photometric catalog. The automated SPOC (Slit Positioning
Optimization Code, Bottini et al. 2005) algorithm is run to maximize
the number of slits given the geometrical and optical constraints of
the VIMOS set-up. We have designed masks with slits of one arcsec
width, and have forced that a minimum of 1.8 arcsec of sky is left on
each side of a targeted object to allow for accurate sky background
fitting and removal during later spectroscopic data processing. On
average, taking advantage of the VIMOS multiplexing capability, a
GOODS pointing of 4 quadrants LR-Blue mask contains up to 360 slits.
Masks with the VIMOS MR grism were designed with a multiplexing of 1
to avoid the superposition of zero and negative orders between
adjacent spectra.  Thus, a pointing of 4 quadrants MR mask can
contain, on average, 150 slits.  Dithering of the targets along the
slits was applied in order to effectively improve the sky subtraction
an the removal of CCD cosmetic defects. In the LR-Blue survey, the
dithering pattern consisted of three position separated by a step of
1.4 arcsec. In the MR survey, the dithering pattern consisted of five
position separated by a step of 1.5 arcsec, in order to provide enough
independent pointings to construct and apply a correction for fringing
at red wavelengths (see section 3.2).

In the LR-Blue campaign, we have used the LR-Blue grism together with
the OS-Blue cutoff filter, which limits the bandpass and order
overlap. With 1 arcsec slits, the resolution is $\sim 28$ $\AA$ and
the dispersion is 5.7 $\AA/\rm{pixel}$. 10 exposures of 24 minutes
each are taken for a total exposure time of 4h per mask. In the MR
campaign, the MR Orange grism is used together with the GG475
filter. With 1 arcsec slits, the resolution is $\sim 13$ $\AA$ and the
dispersion is 2.55 $\AA/\rm{pixel}$. 12 exposures of 20 minutes each
are taken for a total exposure time of 4h per mask. We have asked for
night arc-lamp calibration to reduce problem due to the instrument
instability.

\subsection{Data Reduction} The pipeline processing of the VIMOS-GOODS
data is performed using the VIMOS Interactive Pipeline Graphical
Interface (VIPGI, see Scodeggio et al. 2005 for a full description).
The data reduction is performed in several interactive steps: the
spectra location in the individual spectroscopic frames, the
wavelength calibration, sky subtraction and fringing correction,
combination of the 2D spectra of dithered observations, extraction of
the 1D spectra and flux calibration.  The location of the slits is
known from the mask design process, hence, knowing the grism zero
deviation wavelength and the dispersion curve, the spectra location is
known a priori on the detectors. However, small shifts from predicted
positions are possible due to the complete manufacturing and
observation process. From the predicted position, the location of the
spectra are identified accurately on the 4 detectors and an extraction
window is defined for each slit.  The wavelength calibration is
secured by the observation of night arc-lamps through the observed
mask. Wavelength calibration spectra are extracted at the same
location as the object spectra and calibration lines are identified to
derive the pixel to wavelength mapping for each slit. The wavelength
to detector pixel transformation is fit using a third order
polynomial, resulting in a median rms residual of $\sim 0.7$ $\AA$
across the wavelength range in the LR-Blue masks and $\sim 0.36$ $\AA$
in the MR masks.  A low order polynomial (second order) is fit along
the slit, modeling the sky background contribution at each wavelength
position, and subtracted from the 2D spectrum. In the case of the
LR-Blue campaign the fringing is not present and all 10 exposures of a
sequence are directly combined by shifting the 2D spectra following
the offset pattern to register the object at the same position. The
individual frames are combined with a median, sigma-clipping algorithm
to produce the final summed, sky subtracted 2D spectrum. In the case
of the Medium resolution spectra, the fringing is significant at
$\lambda > 7000$ $\AA$ and needs to be removed. Therefore, a fringing
correction is applied before combining the dithered frames. As the
object is moved to different positions along the slit following the
effect pattern, the median of the 2D sky subtracted spectra produces a
frame from which the object is eliminated, but that includes all
residuals not corrected by sky subtraction, in particular the fringing
pattern varying with position across the slit and wavelength.  This
sky/fringing residuals is, then, subtracted from each individual 2D
sky subtracted frame. The fringing corrected frame are, then, combined
as in the case of the LR-Blue spectra.

The last step done automatically by VIPGI is to extract a 1D spectrum
from the summed 2D spectrum, using an optimal extraction following the
slit profile measured in each slit (Horne, 1986). The 1D spectrum is
flux calibrated using the ADU to absolute flux transformation computed
from the observations of spectrophotometric standard stars.

A final check of the 1D calibrated spectra is performed and the most
discrepant features are removed manually, cleaning each spectrum of
zero order contamination, sky lines residuals and negative non
physical features.

\subsection{The VIMOS LR-Blue wiggles} Spurious wiggles of amplitude
of about 3 to 8\% have been detected in VIMOS MOS spectra taken with
the combination of LR\_Blue grism and OS\_Blue Order Sorting (OS)
filter. The position of the wiggles in the spectrum compares well with
the wiggles in the response curve of the OS\_Blue filter (see also the
ESO VIMOS User Manual, fig. A.3). This clearly indicates that the
wiggles originate in the OS filter. As such, then, the effect of the
wiggles is multiplicative. In principle, spectroscopic screen
flat-fields, even taken during the day, would be sufficient to
correct the wiggles. However, several aspects make this correction
very difficult.  The wiggles position and amplitude are found to
depend on the spectral resolution, which in turn depends on slit width
and object size. The wiggle pattern and the overall shape of the flat
field spectra depend significantly on the position of the slit in the
field of view. In addition, The normalization of flat field spectra is
made problematic by the possible overlap of $0^\mathrm{th}$ order
spectra from neighboring slits. For these reasons we prefer not to
correct the wiggles observed in the LR-Blue spectra.

\begin{figure}
\begin{center}
\begin{minipage}{0.49\textwidth}
\resizebox{\hsize}{!}{\includegraphics{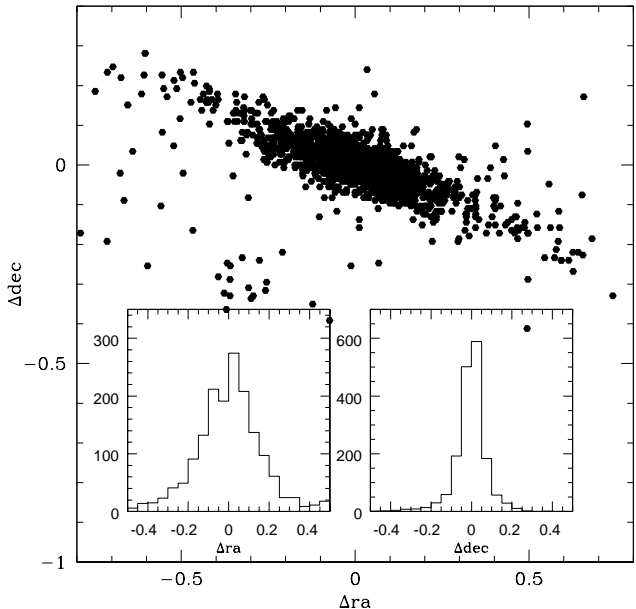}}
\end{minipage}
\end{center}
\caption{ $\Delta ra$ and $\Delta dec$ residual distribution of the
cross-correlation between ``reconstructed'' WFI coordinates and
original target coordinates. The strong distortion of VIMOS quadrants
is not completely removed as indicated by the trend in the $\Delta
ra$-$\Delta dec$ distribution shown in the main panel. However, the
rms of the cross-correlation is lower than 0.2 arcsec in both
coordinates, as shown by the $\Delta ra$ and $\Delta dec$ histograms
in the smaller panels, ensuring an accurate target identification. }
\label{coo}
\end{figure}

\subsection{Target coordinates} Due to the rotation angle of the VIMOS
GOODS pointings (-20 deg), which is different from the default values
accepted by VIPGI (0 and 90 deg), VIPGI does not provide the
astrometry of the extracted spectra. The only information provided by
VIPGI are the coordinates in mm on the focal plane stored in the so
called VIPGI \emph{object table}. To overcome this problem, we
transform the focal plane coordinates of each object into CCD
coordinates trough the appropriate distortion map stored in the header
of our VIMOS observations.  Slits which contain only one object (only
one spectrum extracted) are used to calculate the transformation
matrix from VIMOS to the GOODS R-band WFI CCD coordinates through the
IRAF routines \emph{geomap} and \emph{geoxytran}. As a last step the
WFI $X_{CCD}$ and $Y_{CCD}$ assigned to each extracted spectrum are
converted to $\alpha$, $\delta$ on the basis of the WFI-GOODS
astrometry. These 'reconstructed' coordinates are, then, matched to
the original GOODS VIMOS target catalog to identify the primary
targets and the serendipitous objects. After the identification, the
original WFI target coordinates are assigned to the primary targets
while the secondary objects keep the reconstructed coordinates. Fig.
\ref{coo} shows the distribution of the $\Delta ra$ and $\Delta dec$
in the cross-correlation of the reconstructed WFI coordinates and the
original targets coordinates. The very strong distortion of the VIMOS
CCD is not completely removed as indicated by the trend in the $\Delta
ra$-$\Delta dec$ distribution shown in the main panel. This is due to
the fact the we can use few objects to calculate the transformation
matrix from VIMOS to the GOODS R-band WFI CCD coordinates. However,
the rms of the cross-correlation is lower than 0.2 arcsec in both
coordinates (see the $\Delta ra$ and $\Delta dec$ histograms in the
smaller panels of the same figures), allowing for an accurate target
identification.

It is worth to mention that due to a bug, VIPGI assigns wrong focal
plane coordinates to a small sample of objects in slits with more than
2 spectra. We identify 82 cases in the LR-Blue campaign, of which 80\%
has no redshift determination, and 34 in the MR campaign, of which
50\% has no redshift determination. Those objects have focal plane
coordinates $x_{fp}=0.0,y_{fp}=0$.  Moreover, on the basis of the
reconstructed WFI coordinates these objects are located completely out
of the slit where they should be. The complete list of those objects
is available at http://www.eso.org/science/goods/.

\begin{figure*}
\begin{center}
\begin{minipage}{0.49\textwidth}
 \includegraphics[width=\textwidth]{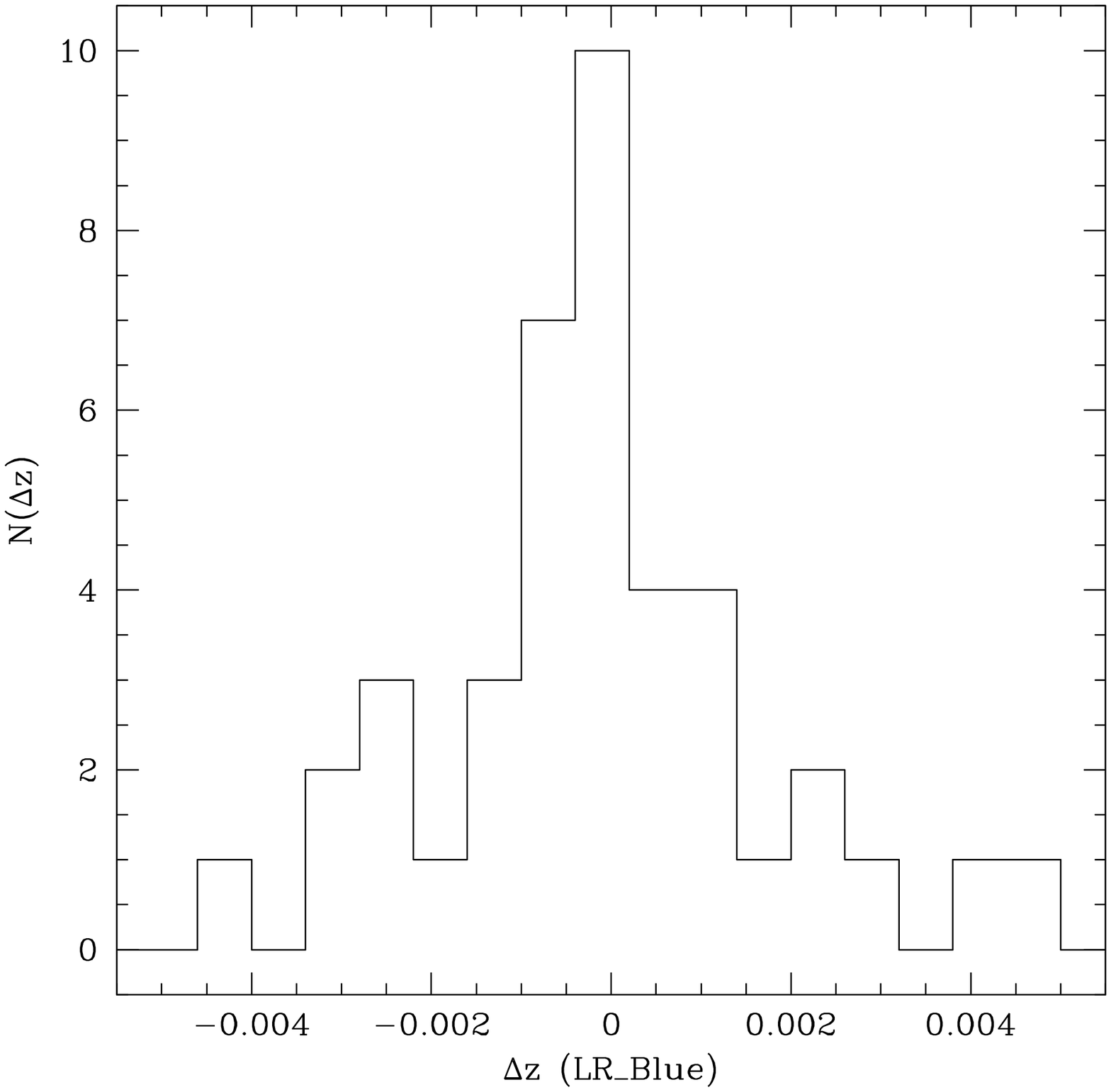}
\end{minipage}
\begin{minipage}{0.49\textwidth}
 \includegraphics[width=\textwidth]{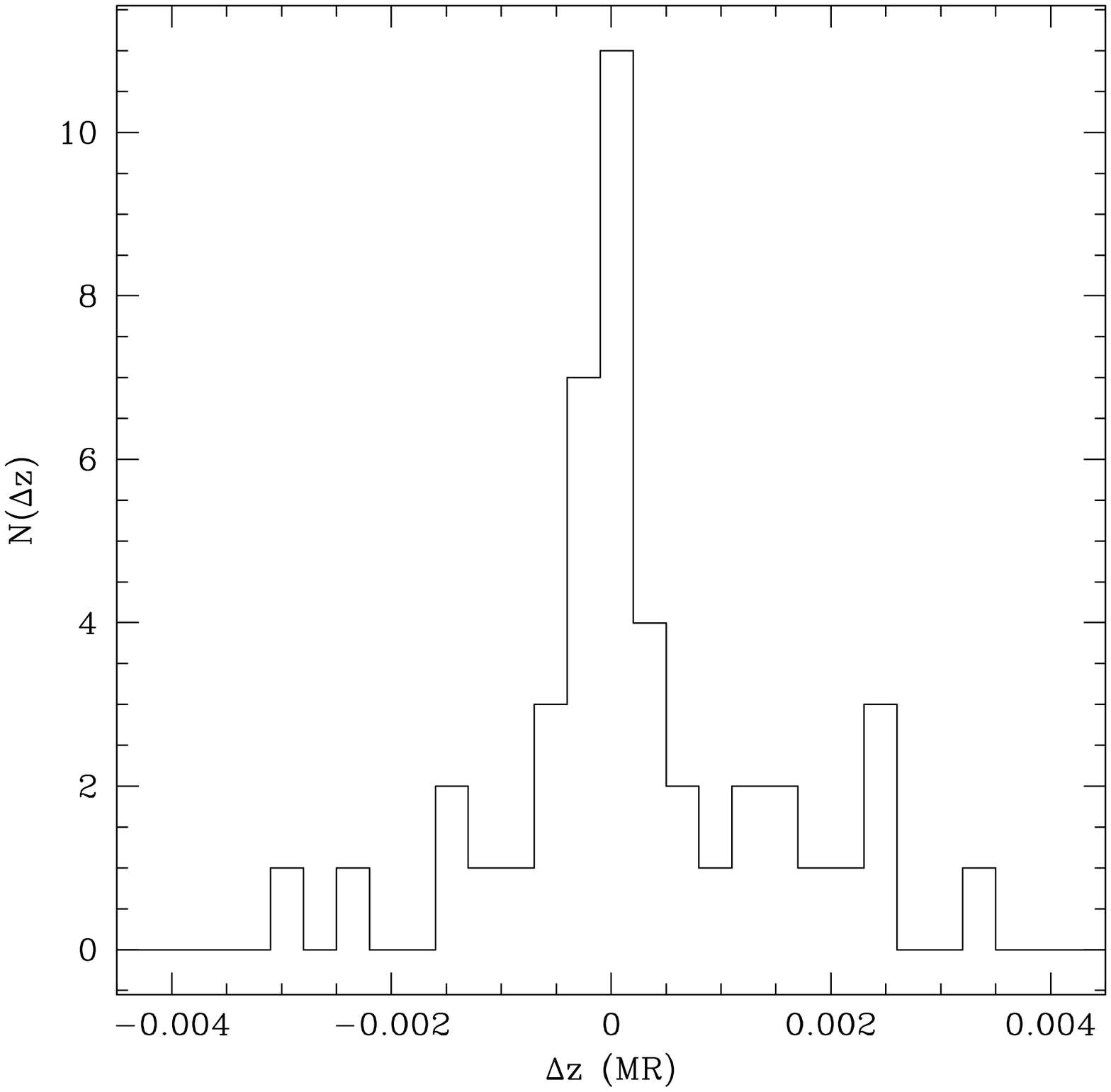}
\end{minipage}
\end{center}
 \caption{Redshift differences between objects observed twice or more
 in independent VIMOS LR-Blue (left panel) and MR (right panel)
 observations. The distribution has a dispersion of $\sigma_z=0.0013$
 in the LR-Blue campaign and $\sigma_{z}=0.0007$ in the MR campaign.}
\label{fig:z_err}
\end{figure*}

\section{Redshift Determination}
2344 spectra  have been extracted from the 6 LR-Blue masks
and 968 have been extracted from 6 MR masks. From them we have been
able to determine 1481 redshifts in the LR-Blue campaign and 656 in
the MR campaign. 33\% of the LR-Blue slits and 18\% of the MR slits
contain more than one spectrum. Most of the secondary spectra obtained
provide additional observations of known targets. We have identified
2235 unique LR-Blue objects and 886 unique MR sources.

In the large majority of the cases the redshift has
been determined through the identification of prominent features of
galaxy spectra: 
\begin{itemize}
\item at low redshift the absorption features: the 4000\AA\ break, Ca H and
K, H$\delta$ and H$\beta$ in absorption, g-band, MgII 2798
\item and the emission features: [O\,{\sc ii}]3727, [O\,{\sc
iii}]5007, H$\beta$, H$\alpha$
\item at high redshift: Ly$\alpha$, in emission and absorption,
ultraviolet absorption features such as [Si\,{\sc ii}]1260, [O\,{\sc
i}]1302, [C\,{\sc ii}]1335, [Si\,{\sc iv}]1393,1402, [S\,{\sc
ii}]1526, [C\,{\sc iv}]1548,1550, [Fe\,{\sc ii}]1608 and [Al\,{\sc
iii}]1670
\end{itemize} 

The redshift estimation has been performed cross-correlating the
individual observed spectra with templates of different spectral
types: S0, Sa, Sb, Sc, Ell. for the low redshift spectra, several
Lyman break galaxies templates in absorption and emission, BzK and AGN
template. The cross-correlation is performed by using the \emph{rvsao}
package (\emph{xcsao} routine) in the IRAF environment on each
spectrum. In particular, an error and trial approach is used for the
$z > 1.8$ galaxies, whose redshift determination is made difficult by
the low S/N spectral absorption features and the wiggles in the
LR-Blue spectrum.

In analogy to the complementary GOODS-FORS2 redshift campaign, we use
four quality flags to indicate the quality of each redshift
estimation. The determination of the quality flag is done in two
steps. As a first step, the assignment of the quality flag is done
during the cross-correlation of the spectrum with the templates on the
basis of the cross-correlation coefficient provided by the routine
\emph{xcsao} in the IRAF environment. The quality flags are assigned
with the following criteria:
\begin{itemize}
\item flag A: high quality, values of the \emph{xcsao} correlation
coefficient $R \ge 5$; emission lines and strong absorption features
are well identified.
\item flag B: intermediate quality, values of the \emph{xcsao} correlation
coefficient $3 \le R < 5$; one emission line plus few absorption
features are well identified.
\item flag C: low quality, values of the \emph{xcsao} correlation coefficient
$R < 3 $, features of the continuum not well identified.
\item flag X: no redshift estimated, no features identified.
\end{itemize}

As a second step, each spectrum with the superposed main spectral
features, is checked by eye by different people and a refinement of the
redshift determination and the quality flag assignment is
performed. On average, each spectrum is checked more then three times.

In $\sim 15\%$ of the cases the redshift is based only on one emission
line, usually identified with [O\,{\sc ii}]3727 or Ly$\alpha$.  In
these cases the continuum shape, the presence of breaks, the absence
of other spectral features in the observed spectral range and the
broad band photometry are particularly important in the evaluation.
In general these solo-emission line redshifts are classified as
``likely'' (B) or ``tentative'' (C) if the no other information are
provided by the continuum.

The internal redshift accuracy can be estimated from a sample of
galaxies which have been observed twice in independent VIMOS mask
sets.  We find 39 of such objects in the LR-Blue masks and 40 in the
MR masks with quality flag A and B. $\sim 45\%$ of this objects have
been observed as serendipitous objects. The distribution of measured
redshift differences is presented in Figure~\ref{fig:z_err}. The mean
of the LR-Blue (left panel of the figure) and the MR (right panel of
the figure) $\Delta z$ distribution is close to zero ($\sim 10^{-5}$)
in both cases.  The redshift dispersion is $\sigma_z=0.0013$ ($\sim
400$ $\rm{kms}^{-1}$) for the LR-Blue objects and $\sigma_z=0.0007$
($\sim 200$ $\rm{kms}^{-1}$) for the MR redshifts. This latter
estimation is in very good agreement with the value obtained in the
GOODS-FORS2 survey (Vanzella et al. 2005), conducted on similar
objects using similar spectral resolution and spectral range as VIMOS
MR. We note that the mean values of the redshift estimation
uncertainty estimated in this way are $\sim 3$ times larger than the
mean error ($\sigma_z=0.0004$ in the LR-Blue survey and
$\sigma_z=0.0002$ in the MR survey) calculated by the IRAF routine
\emph{xcsao}.

\begin{table}[h]
\caption{Success rate of the GOODS VIMOS LR-Blue campaign. The first
column lists the name of the target family, the second column lists
the fraction of the target catalog due to the corresponding color
selection (BzK and sub-udrpout family overlap largely but they are
considered as separated family in the table). The third column lists
the success rate (fraction A$+$B flag objects) of each target
family. The last four columns list the percentage of A, B, C and X
flag redshift determinations, respectively.}
\begin{center}
\begin{tabular}[h]{|c|c|c|c|c|c|c|}
\hline
Target & fraction & s.r. & A & B & C & X \\
\hline
\hline
U-dropouts & 16\% & 49\% & 35\% & 14\% & 15\% & 36\% \\
\hline
BzK        & 51\% & 36\% & 24\% & 12\% & 21\% & 43\% \\
\hline
sub-U-dropouts  & 66\% & 51\% & 35\% & 16\% & 16\% & 33\% \\
\hline
X-ray      &  5\% & 51\% & 37\% & 14\% & 14\% & 35\% \\
\hline
\end{tabular}
\end{center}
\label{tab}							   
\end{table}

\subsection{The success rate}

We measured a redshift for 63\% (70\% for the primary targets) of the
observed LR\_Blue spectra. However, to estimate the success rate of
the surveys we use only the objects with high quality flags, A and
B. In the LR-Blue survey the success rate is 39\% if we consider the
whole sample and 48\% if we consider only the primary targets. The
serendipitous sources, which account for 33\% of the sample, are
usually faint neighbor and lie often at the edge of the 2D
spectrum. For these objects the success rate is very low, $\sim
20\%$. We have investigated how the success rate depends on the target
selection and on the redshift windows. As described in Sec. 2.2, the
LR-Blue targets can be divided in 4 families: U-dropouts, BzK objects,
sub-udroputs and X-ray sources. BzK and sub-udroputs samples largely
overlap. In this discussion we consider the two family
individually. Table \ref{tab} shows the fraction of targets observed
in the analyzed masks and the corresponding success rate for each
target family. The BzK objects have the lowest success rate. 64\% of
those objects have flag C or no redshift (flag X) leading to a very
low succes rate. All the other target families have a success rate of
$\sim 51\%$. In addition we find a strong dependence of the success
rate on the redshift window. In particular:
\begin{itemize}
\item the objects at low redshift ($z_{spec} <
1$) and in the range $2.2 \le z_{spec} \le 3.5$ have the highest
fraction of A flags, $\sim 60\%$
\item the very high redshift galaxies ($z_{spec} > 3.5$) have mainly
quality flag B because the Ly$\alpha$ in emission is in the very
red part of the spectrum and the other features are not well
identified
\item the objects at $1.8 < z_{spec} < 2.2$ show the highest fraction
of insecure redshift determinations (percentage of the C flag
determination $\sim 65\%$)
\end{itemize}

There are two main reasons for the high failure rate at redshift $1.8 <
z_{spec} < 2.2$ and for the BzK galaxies, which lie mainly in this
redshift range . The first one is that the Ly$\alpha$ is generally
outside the spectral range covered by the LR-Blue grism, at $\lambda <
3600 \AA$. Therefore, other 'secondary' spectral features, such as
SII, OI, CII, SiIV, SII, and CIV, have to be used to estimate the
redshift. However, these features are not as strong as the Ly$\alpha$
in absorption or emission. In addition, at $\lambda < 4000 \AA$ the
VIMOS efficiency drops very quickly and the presence of the wiggles
described in Sec.3.3, makes the redshift determination very
insecure. In fact, only for the very bright sources the S/N at
$\lambda < 4000-4200$~\AA\ is high enough to identify the 'secondary'
features of the continuum.

We measured a redshift for 67\% (75\% for the primary targets) of the
observed MR spectra. In the VIMOS MR campaign the overall success rate
(A+B flag redshifts) is 60\% and reaches the 65\% level if only the
primary targets are considered. We do not note any dependence on the
target selection criteria and redshift windows.

\section{Discussion}

\subsection{Reliability of the redshifts - comparison with previous surveys}

A practical way to assess the reliability of the redshifts is to
compare the present results with independent measurements of other
surveys.  For this purpose we use the results of four surveys
conducted on the same field: the GOODS-FORS2 campaign (Vanzella et
al. 2005, 2006, 2008), the K20 survey (Cimatti et al. 2002), Szokoly
et al. (2004) and the VVDS survey (Le Fevre et al.  2005).  To create
a secure redshift reference sample, we have selected only the high
quality redshift determinations of those surveys: GOODS FORS2 quality
A and B, K20 quality 1, VVDS quality 3 and 4 and Szokoly et al. 2004
quality 3 and 2$+$ redshifts, which have all a confidence higher than
95\%).

Among the LR-Blue redshift determinations, there are 113 VIMOS objects
in common with this high quality reference sample within a spatial
accuracy of 0.5 arcsec or better. 58 of them have VIMOS quality flag
A, 16 have flag B, 16 have flag C and 23 do not have a redshift
estimation (flag X). These 23 objects lie in a redshift range, $0.8 <
z < 1.7$, not accessible to VIMOS LR\_Blue. 27 cases of the A,B and C
quality redshifts show ``catastrophic'' discrepancies
($|z_{VIMOS}-z_{FORS2/k20/VVDS}| > 0.015$), which account for 5 of the
VIMOS flag A objects, 8 of the flag B and 11 of the flag C sources. 

After visual comparison of the VIMOS
and FORS2/K20/VVDS/CDF spectra we find that 3 of the 5 VIMOS flag A
spectra showing ``catastrophic'' discrepancies are wrong VIMOS
redshift determinations. 

\begin{itemize}
\item[-] VIMOS GOODS\_LRb\_001\_q2\_1\_1 versus FORS2
GDS\_J033217.78-274823.8: the [OIII] in emission is identified in the
FORS2 spectrum and it is hidden by a strong sky line residual in the
VIMOS spectrum. Thus, the [OII] in the VIMOS spectrum is missclassified
as Ly$\alpha$

\item[-]VIMOS GOODS\_LRb\_001\_1\_q1\_51\_1 versus FORS2
GDS\_J033226.67-274013.4: the [OII] in the FORS2 spectrum is identified
at z=1.612, a redshift window not accessible to VIMOS LR-Blue. No
emission lines are visible in the VIMOS spectrum and the low S/N UV
absorption features are missclassified.

\item[-]VIMOS GOODS\_LRb\_001\_q2\_35\_1 versus VIMOS LR-Red VVDS 32126:
the strong UV absorption features identified in the VIMOS spectrum
provide a $xcsao$ correlation coefficient similar to that of the FeII
and NeV absorption features identified in the VVDS spectrum. We have
combined the two spectra and re-performed the cross correlation. The
highest correlation peak corresponds to the VVDS redshift estimation.

\item[-]VIMOS GOODS\_LRb\_001\_q3\_71\_2 versus VIMOS LR-Red VVDS
  16975: the VIMOS LR-Blue source is an emission line galaxy and the
  reference VVDS spectrum is clearly an early type galaxy without any
  emission line. The two spectra can not refer to the same
  object. Since 16975 is a secondary object and not a primary target,
  we suspect that the coordinates provided by VIPGI (used to reduce
  the VVDS data) could be wrong as explained in Sec. 3.4. Thus, we
  consider the VIMOS redshift estimation correct.

\item[-]VIMOS GOODS\_LRb\_002\_q2\_55\_1 versus FORS2
GDS\_J033221.94-274338.8: the strong emission line in the VIMOS
spectrum is identified as a Ly$\alpha$ due to the absence of
$\rm{H}_{\beta}$ and [OIII] emissions and due to the photometry (the
target is a U-drop-out). If something, the emission could be
classified as a [OII] at much lower redshift (z=0.166) with a much
lower $xcsao$ correlation coefficient. In both cases the FORS2
redshift is not in agreement. We have combined the two spectra and
re-measured the redshift. The correlation gives a good result only
with a Ly$\alpha$ emitter template at z=2.576. No match is found for
the emission seen in the FORS2 spectrum which has a very low S/N. We
think that the line identified as [OII] in the FORS2 spectrum is
instead due to a fringing residual since it is sitting on a sky
line. Thus, the consider the VIMOS redshift estimation correct.
\end{itemize}

The resulting confidence level in the flag A redshift
determinations is 95\% (3 mistakes out of 58 redshift
determinations). Among the flag B spectra showing ``catastrophic''
discrepancies, 6 VIMOS redshift determinations are wrong mainly due to
the presence of sky residuals and $0^{th}$ order of neighboring
spectra, and only 2 are more convincing than the FORS2/K20/VVDS/CDF
determinations. The resulting confidence level is 62\% (6 mistakes out
of 16 redshift determinations). In all the Flag C discrepancies cases
the FORS2/K20/VVDS/CDF redshift determinations are more convincing
than the LR-Blue identified ultraviolet features. The resulting
confidence level is 31\%. However, it is important to note that the
VIMOS flag C is assigned in the majority of the cases to redshifts in
the range $1.8 < z < 2.2$ as explained in Sec. 4.1. The redshift
surveys considered in this comparison (FORS2, K20, VVDS and CDF) are
not designed to cover this redshift range. Thus, such comparison can
only reveal the mistakes and can not provide confirmations to our
estimates. In conclusion, 19 redshift determinations out of 90 are
wrong resulting in an overall confidence level of 78\% in the LR-Blue
VIMOS redshifts. For the 71 cases out of 90 which show good agreement,
we find a mean difference $<z_{LR-Blue-VIMOS}-z_{FORS2/k20/VVDS}>$ =
0.0018 $\pm$ 0.0019, which confirms the mean uncertainty $\Delta z$
found in Sec. 4.

The comparison between the VIMOS MR redshift determinations and
FORS2/K20/VVDS/CDF values is simplified by the fact that the MR survey
is very similar in design to the considered surveys. There are 94
VIMOS objects in common with the high quality reference sample within
a spatial accuracy of 0.5 arcsec. 69 of them have VIMOS quality flag
A, 17 have quality flag B and 8 have quality flag C. We find 5
``catastrophic'' discrepancies: 1 of them have flag A, 1 have flag B
and 3 have flag C:

\begin{itemize}
\item[-] flag A case GOODS\_MR\_new\_1\_d\_q3\_22\_1 versus FORS2
  GDS\_J033243.19-275034.9: an accurate analysis is provided by
  Vanzella et al. (2006, see their Figure 2).  The continuum shows
  increasing bumps/bands in the red, very similar to typical cold
  stars.  After visual inspection of the ACS color image Vanzella et
  al. 2006 concluded that GDS\_J033243.19-275034.9 is a simultaneous
  spectrum of two very close sources: a star and a possible high-z
  galaxy.

\item[-]B flag case VIMOS GOODS\_MR\_new\_1\_d\_q3\_22\_1 versus FORS2
GDS\_J033249.04-2705015.5: the spectral features used for the
identification are all at $\lambda > 7500$, where the fringing is
very strong. The corresponding FORS2 spectra, which suffer less of
fringing, show more convincing spectral features.

\end{itemize}

In the 3 flag C cases, the FORS2/K20/VVDS/CDF redshift estimates seem
to be more robust than the VIMOS redshifts. In all three cases the
spectral features used to identify the redshift are in the region
strongly affected by fringing.

Thus, we obtain a confidence level of 98\% for the quality A MR
redshifts (1 mistake out of 69 redshifts), 94\% for the quality B
redshifts (1 mistakes out of 17 determinations) and 62\% for the
quality C cases (3 mistakes out of 8 determinations).  The overall
confidence level of the redshift determinations of the MR redshift
survey is 95\%.  For the 89 cases out of 94 which show good agreement,
we find a mean difference $<z_{MR-VIMOS}-z_{FORS2/k20/VVDS}>$ = 0.0013
$\pm$ 0.0012.

\begin{figure}
\begin{center}
\begin{minipage}{0.49\textwidth}
\resizebox{\hsize}{!}{\includegraphics{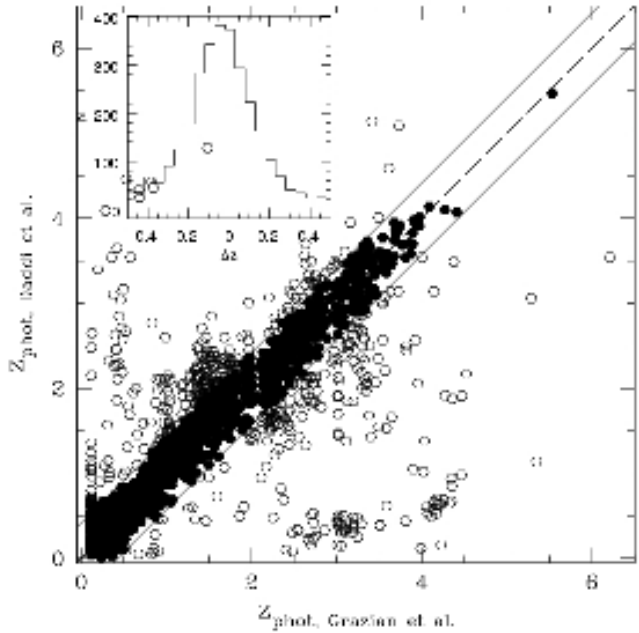}}
\end{minipage}
\end{center}
\caption{Comparison of the GOODS-MUSIC and the GOODS photometric
redshift catalog. The dashed line shows the slope 1 line and the solid
lines show the 3$\sigma$ limits. The small panel within the main frame
shows the $\Delta z=z_{MUSIC}-z_{GOODS}$ distribution with
$\sigma=0.13$. }
\label{zgzd}
\end{figure}

\begin{figure*}
\begin{center}
\begin{minipage}{1.0\textwidth}
\resizebox{\hsize}{!}{\includegraphics{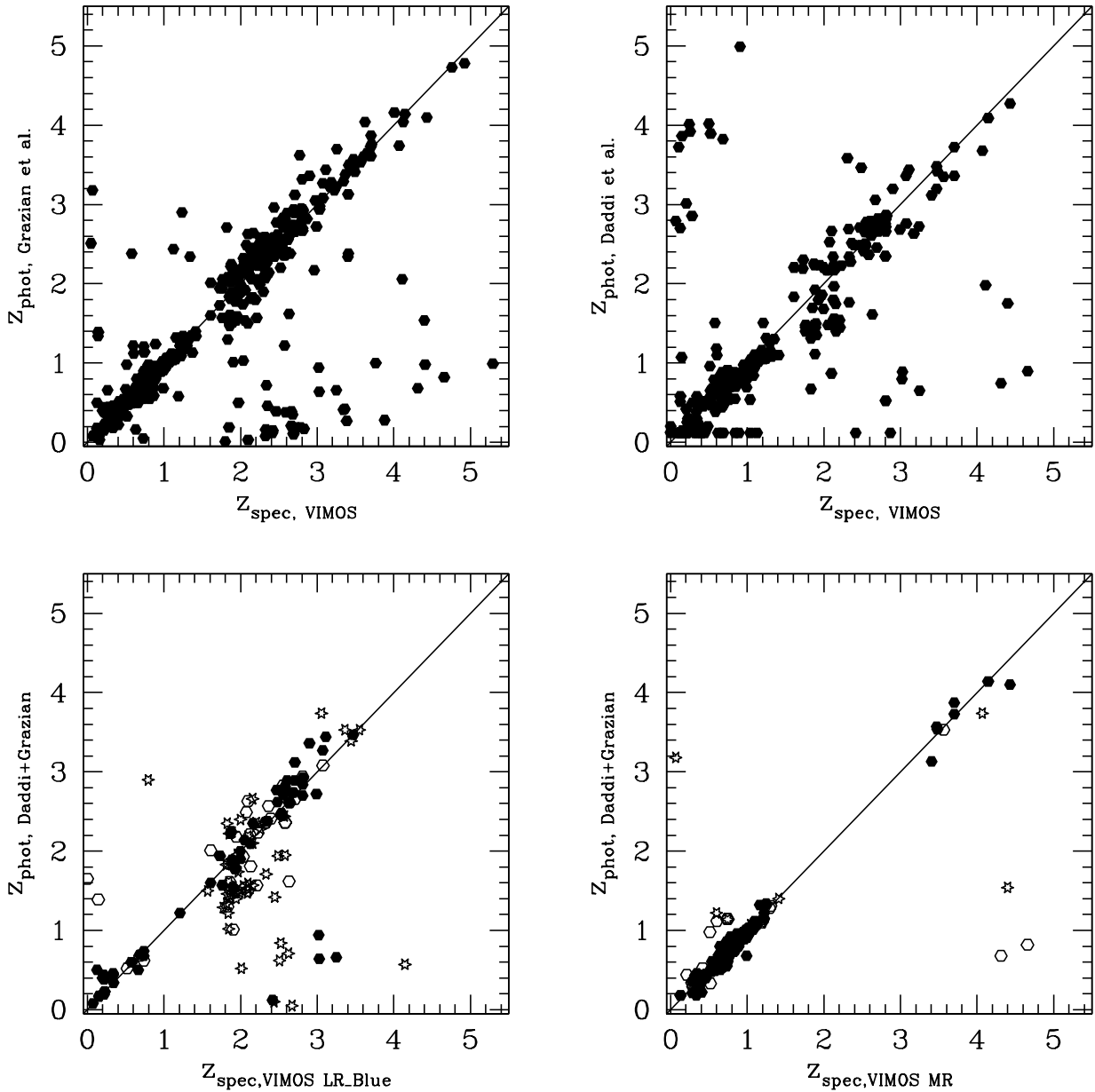}}
\end{minipage}
\end{center}
\caption{$z_{spec}$ versus $z_{phot}$. The two top panels show the
comparison between the GOODS VIMOS (LR-Blue + MR) high quality
$z_{spec}$ and the GOODS-MUSIC $z_{phot}$ of Grazian et al. 2006 (top
left panel) and the GOODS $z_{phot}$ of Daddi et al. (private
communication, top right panel). In both cases there is a rather high
percentage of discrepancies located in different regions of the
diagram. If we compare the LR-Blue (bottom left panel) and the MR (the
bottom right panel) $z_{spec}$ only with $z_{phot}$, which are
consistent within $3\sigma$ in the GOODS-MUSIC and GOODS catalogs, the
agreement is much higher. In both bottom panels the filled circles are
flag A $z_{spec}$, the empty circles are B flag $z_{spec}$ and the
stars are C flag $z_{spec}$.}
\label{zs_zphot}
\end{figure*}

\begin{figure*}
\begin{center}
\begin{minipage}{0.33\textwidth}
 \includegraphics[width=\textwidth]{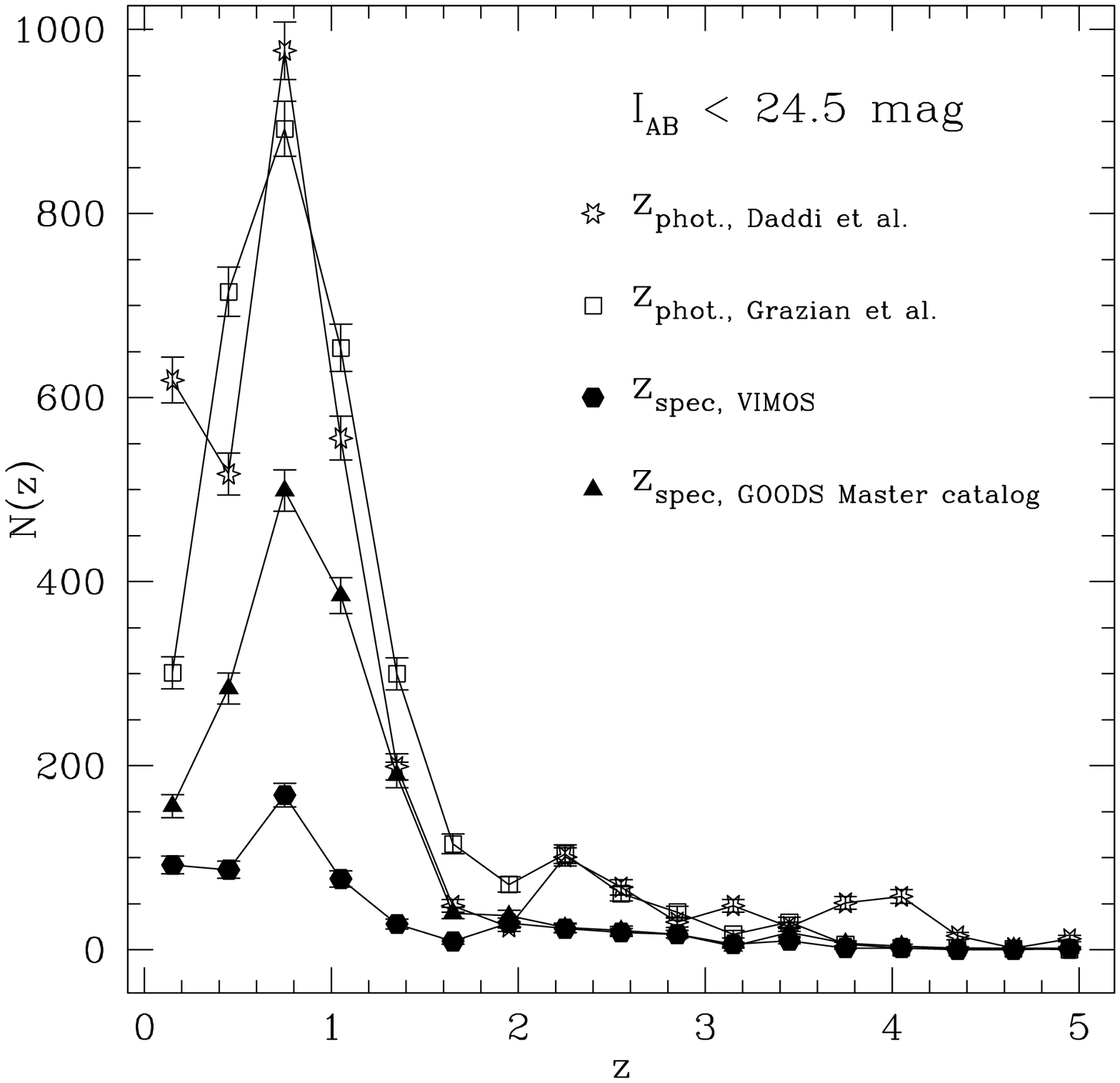}
\end{minipage}
\begin{minipage}{0.33\textwidth}
 \includegraphics[width=\textwidth]{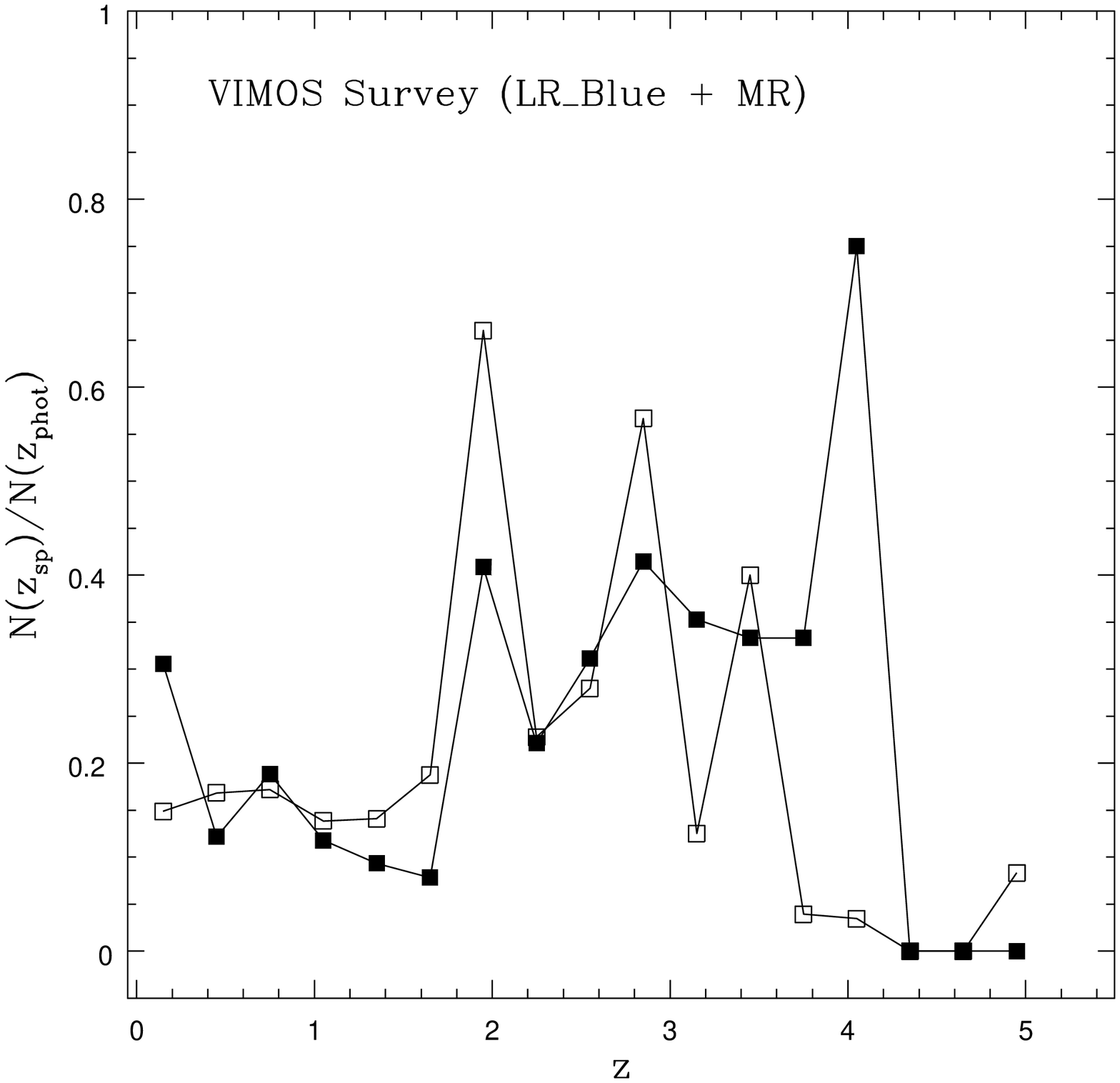}
\end{minipage}
\begin{minipage}{0.33\textwidth}
 \includegraphics[width=\textwidth]{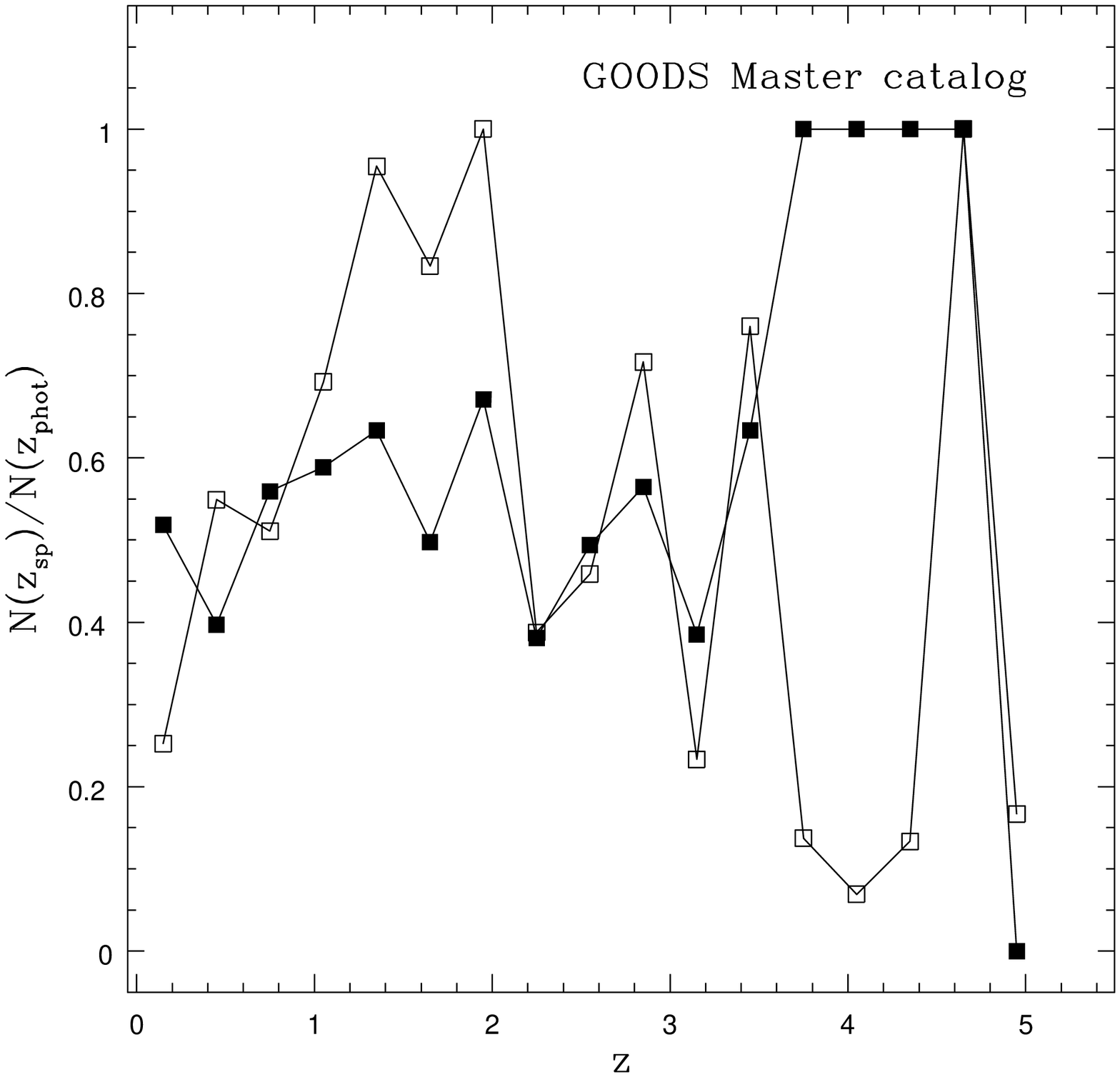}
\end{minipage}
\end{center}
 \caption{Completeness level of the GOODS survey in the CDF-S. The
 left panel shows the coarse-grain redshift distribution of the
 $z_{phot}$ catalogs considered for the comparison and the
 spectroscopic catalogs (GOODS VIMOS catalog and the GOODS master
 catalog containing all the available spectroscopic redshifts in the
 CDF-S). The central and the right panels show the spectroscopic
 completeness level of the GOODS VIMOS and the master catalogs,
 respectively, in several redshift bins. The completeness level is
 defined as the ratio $N(z_{spec})/N(z_{phot})$ in each redshift
 bin. In both panels the empty squares show the
 $N(z_{spec})/N(z_{phot})$ obtained from the GOODS-MUSIC catalog of
 Grazian et al.(2006) and the filled squares show the
 $N(z_{spec})/N(z_{phot})$ obtained from the GOODS catalog of Daddi et
 al.(private communication).}
\label{cp}
\end{figure*}

\begin{figure}
\begin{center}
\begin{minipage}{0.4\textwidth}
 \includegraphics[width=\textwidth]{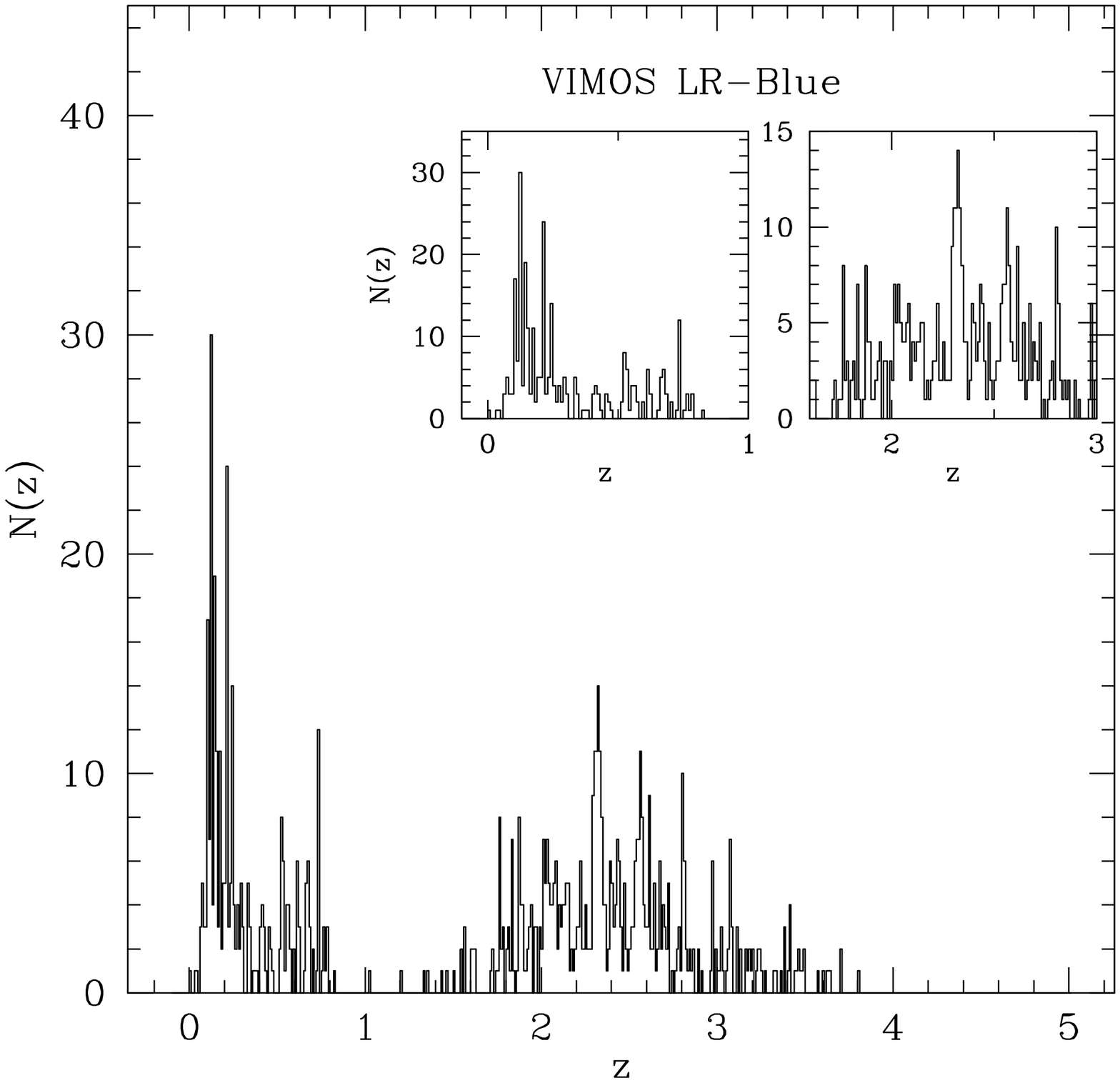}
\end{minipage}
\begin{minipage}{0.4\textwidth}
 \includegraphics[width=\textwidth]{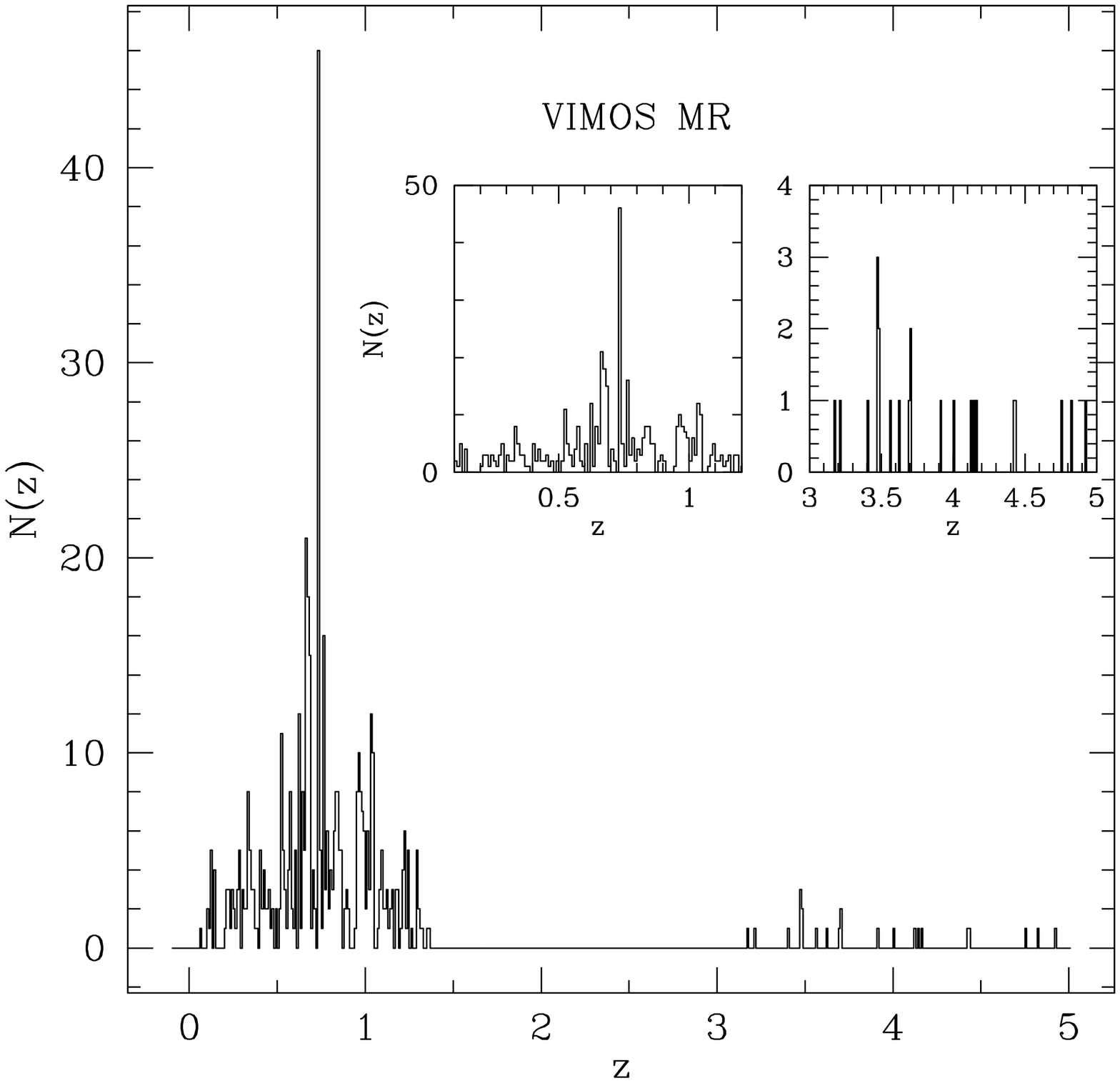}
\end{minipage}
\begin{minipage}{0.4\textwidth}
 \includegraphics[width=\textwidth]{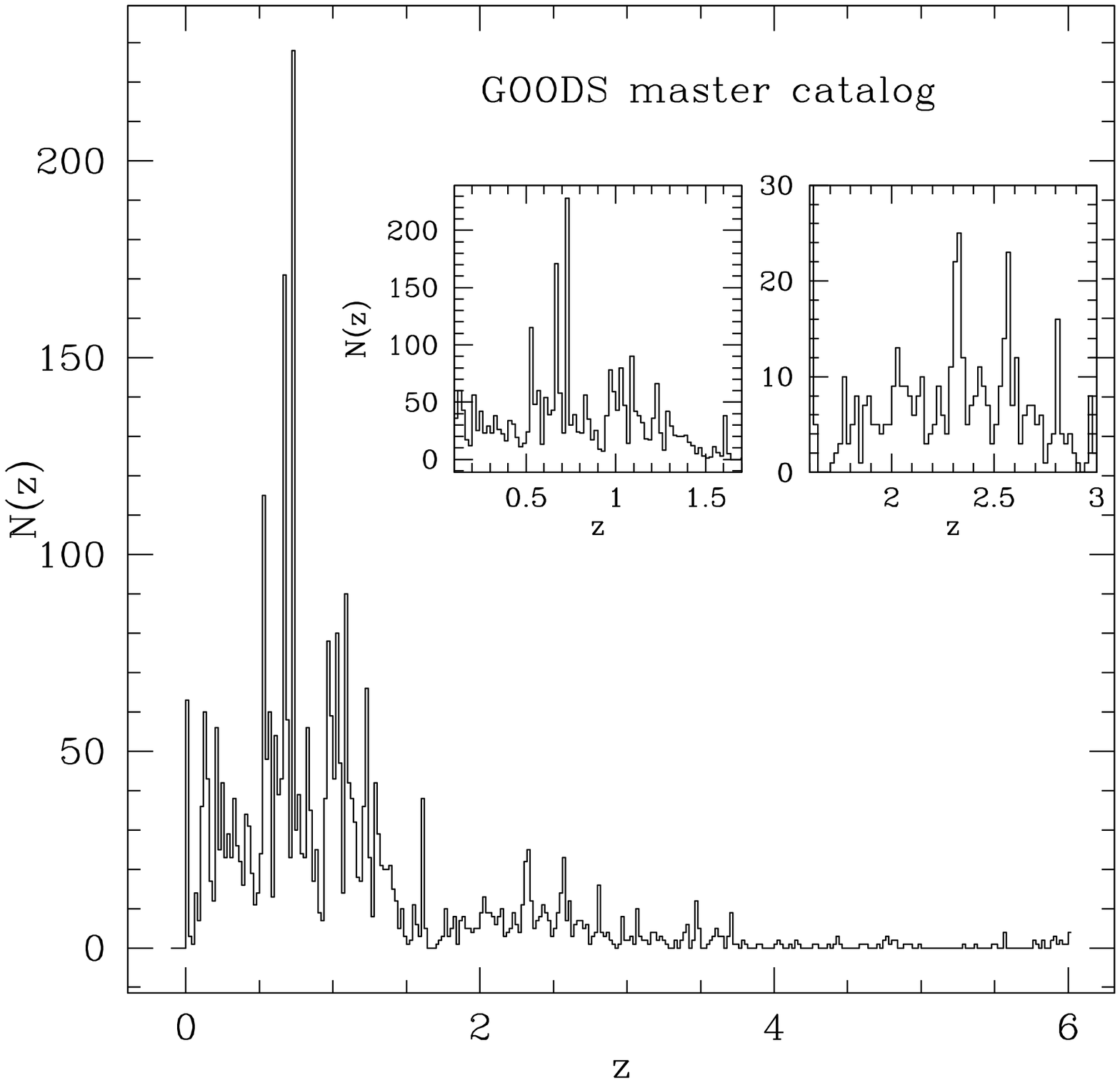}
\end{minipage}
\end{center}
 \caption{Fine-grain redshift distribution of the available
 spectroscopic catalogs: the VIMOS LR-Blue catalog in the top
 panel, the VIMOS MR catalog in the central panel and the GOODS master
 catalog in the bottom panel. The smaller panels within the main frames
 show the distribution in the redshift region of particular interest.}
\label{zdist}
\end{figure}

\subsection{Reliability of the redshifts - comparison with photometric redshift}
An alternative way to assess the reliability of the redshifts is to
compare the present results with accurate photometric
redshifts. Photometric redshift determinations are inevitably plagued
by a rather high incidence of catastrophic failures and present biases
depending on the redshift determination procedure applied. Thus, to
overcome these problems, we use simultaneously two different
photometric redshift catalogs: the GOODS-MUSIC catalog (Grazian et
al. 2006) and a GOODS photometric redshift catalog (Daddi et al.,
private communication). The GOODS-MUSIC photometric redshifts are
based on a high quality multiwavelength (from 0.3 to 8.0 $\mu$mm)
catalog, which includes accurate 'FSF-matched' ACS, $JHKs$ ESO VLT,
Spitzer IRAC and the first 3 h U-band VLT-VIMOS magnitudes. It is
trained on the high quality GOODS-FORS2 and VVDS spectroscopic
redshifts. The Daddi et al. (private communication) catalog is defined
using a GOODS multicolor catalog including IRAC but with the exclusion
of the VIMOS U-band, and trained using a whole set of high quality
GOODS-FORS2, K20 and GMASS redshifts. For our purposes we have
cross-correlated the two photometric redshift catalogs and created a
high quality reference sample which includes only the objects with
concordant GOODS-MUSIC and GOODS redshift estimations. We have
calculated the $\sigma$ of the $\Delta z=z_{Grazian}-z_{Daddi}$
distribution and identified the reference objects as those showing a
$|\Delta z| < 3\sigma$. Fig. \ref{zgzd} shows the $z_{Grazian}$ versus
$z_{Daddi}$. The filled circles lying within the $3\sigma$ lines (the
solid lines in the figure) are those included in the reference
photometric redshift sample and the empty circles are excluded from
it.  We have, then, compared this high quality reference $z_{phot}$
sample with the VIMOS LR-Blue and MR spectroscopic
catalog. Fig. \ref{zs_zphot} shows the result of the comparison. We
define as ``catastrophic'' discrepancies the redshift determination
with $|z_{spec}-z_{phot}| > 3\sigma$. We list below the results
obtained for the LR-Blue survey:
\begin{itemize}
\item we find 150 common objects between the $z_{phot}$ reference
sample and the LR-Blue spectroscopic catalog. 65 of them have flag A,
34 have flag B and 51 have flag C
\item there are 4 flag A ``catastrophic'' discrepancies: 1 of them is
a secure Lyman break galaxies with strong Ly$\alpha$ in emission and
well identified ultraviolet features and is not consistent with the
$z_{phot}=0.94$. In the remaining 3 spectra the emission line is
identified as Ly$\alpha$ but it could be also an [OII] as suggested by
the $z_{phot}$. Thus 3 $z_{spec}$ determinations out of 65 can be
considered wrong, which confirms a confidence level of 95\% in the low
resolution flag A redshifts.
\item we find 8 flag B discrepancies: 2 of them are secure low
redshift emission line galaxies ([OII], H$\beta$ and [OIII] well
identified). The remaining 6 spectra are solo-emission line ([OII] or
Ly$\alpha$) spectra with few other low S/N features identified. If the
line is identified differently ([OII] instead of Ly$\alpha$ or
vice-versa) the resulting $z_{spec}$ is consistent with
$z_{phot}$. Thus, we consider these measurements wrong. The
resulting confidence level is 82\%
\item there are 23 flag C $z_{spec}$ which are not confirmed by the
$z_{phot}$, which results in a confidence level about 55\%
\end{itemize}

We list below the results obtained for the MR survey:
\begin{itemize}
\item we find 177 common objects between the $z_{phot}$ reference
sample and the MR spectroscopic catalog. 123 of them have flag A,
37 have flag B and 17 have flag C
\item there are 2 flag A ``catastrophic'' discrepancies. Both cases
are secure low redshift emission line galaxies ([OII], H$\beta$ and [OIII]
well identified). Thus the confidence level of the MR A flag redshifts
is confirmed to be 100\%
\item we find 6 flag B discrepancies: 2 of them are secure low
redshift emission line galaxies. In the remaining 4 spectra, the
emission line is located in the fringing region and could be
misclassified. The resulting confidence level is 89\%
\item there are 4 flag C $z_{spec}$ which are not confirmed by the
$z_{phot}$, which results in a confidence level about 76\%
\end{itemize}

\begin{figure}
\begin{center}
\begin{minipage}{0.49\textwidth}
\resizebox{\hsize}{!}{\includegraphics{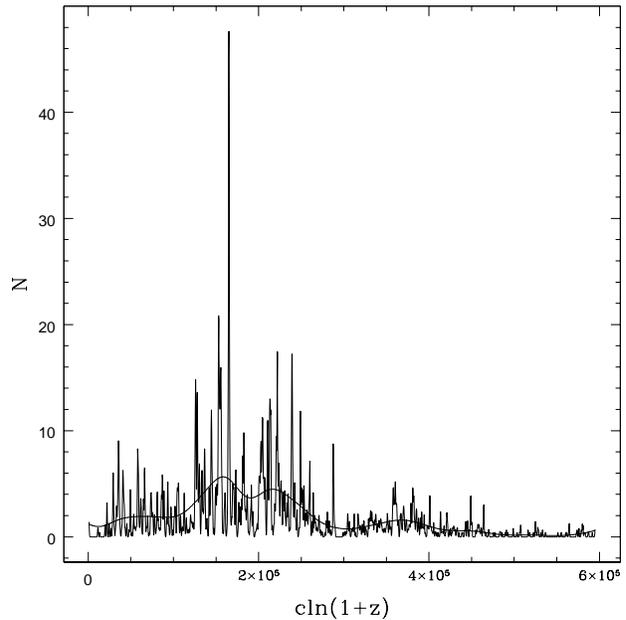}}
\end{minipage}
\end{center}
\caption{Galaxy density in velocity space. The solid line is the
background distribution obtained by smoothing the observed
distribution with a Gaussian with $\sigma_C=15000$ km$s^{-1}$ (solid
line). The galaxy distribution is recomputed using $\sigma_S=300$
km$s^{-1}$ (histogram).}
\label{lss}
\end{figure}

\subsection{The survey completeness}
The main purpose of the two complementary GOODS-South redshift
surveys, the FORS2 and VIMOS campaigns, is to provide a highly
complete spectroscopic sample down to $i_{775}=25$ mag. Thus, it is
very important to know which is the real level of completeness reached
so far after the completion of the whole FORS2 survey and 60\% of the
VIMOS survey. In principle, the selection function of a spectroscopic
survey could be estimated by to the comparison with appropriate
simulations able to reproduce the results of the applied target
selection criteria. In the case of the FORS2 and VIMOS campaigns this
is complicated by the fact that the selection criteria are not uniform
throughout the survey. In fact, they have been tailored for each
observing run in order to optimizing the survey success rate in terms
of redshift estimation on the basis of partial results. To overcome
this problem we use here a different approach. We compare our
spectroscopic redshift catalog with a fairly complete photometric
redshift catalog. As in the previous section we use two $z_{phot}$
catalogs, the GOODS-MUSIC catalog and the GOODS catalog of Daddi et
al. to take under control possible biases. The largest fraction of the
GOODS-MUSIC sample is 90\% complete at $z\sim26$ or $Ks \sim 23.8$ mag
(AB scale). In a similar way, the GOODS catalog of Daddi et
al. includes all the GOODS sources with $Ks <22$ mag (Vega
scale). Since we are calculating the completeness level of our
spectroscopic catalog in the ACS i band, we have checked that both
$z_{phot}$ catalogs are able to reproduce within the errors the
observed number counts in the considered band down to the required
magnitude limit ($i_{775}=25$). As shown in the left panel of
Fig. \ref{cp}, both $z_{phot}$ catalogs are able to reproduce the same
coarse-grain redshift distribution within $3\sigma$. The redshift bin
is chosen to be $\delta z=0.3$ very close to the $3\sigma$ uncertainty
obtained in the comparison of the two $z_{phot}$ catalogs. The same
panel shows also the coarse-grain redshift distribution of the VIMOS
(LR-Blue+MR, the filled circles) survey and of the GOODS ``master
catalog'', namely the compilation of all high quality spectroscopic
redshifts (GOODS FORS2 quality A and B, VIMOS LR-Blue+MR quality A and
B, K20 quality 1, VVDS quality 3 and 4 and Szokoly et al. 2004 quality
3 and 2$+$ redshifts), cleaned of double observations. The
completeness level in each redshift bin is calculated as the ratio
$N(z_{spec}/N(z_{phot})$. The central panel of Fig. \ref{cp} shows the
redshift-dependent completeness level of the whole GOODS-VIMOS survey,
and the right panel shows the same results for the GOODS master
catalog. The two $z_{phot}$ catalogs provide consistent results within
the error bars in both panels till redshift $z=3.5$. The completeness
level of the GOODS master catalog is rather high, $\sim 60\%$ till $z
\sim 3.5$.

At higher redshift the discrepancy is larger than $3\sigma$ (the error
bars are not shown in the central and right panels for clarity). This
large discrepancy do not allows to draw any conclusion in this
redshift range.

\begin{figure*}
\begin{center}
\begin{minipage}{0.36\textwidth}
\resizebox{\hsize}{!}{\includegraphics{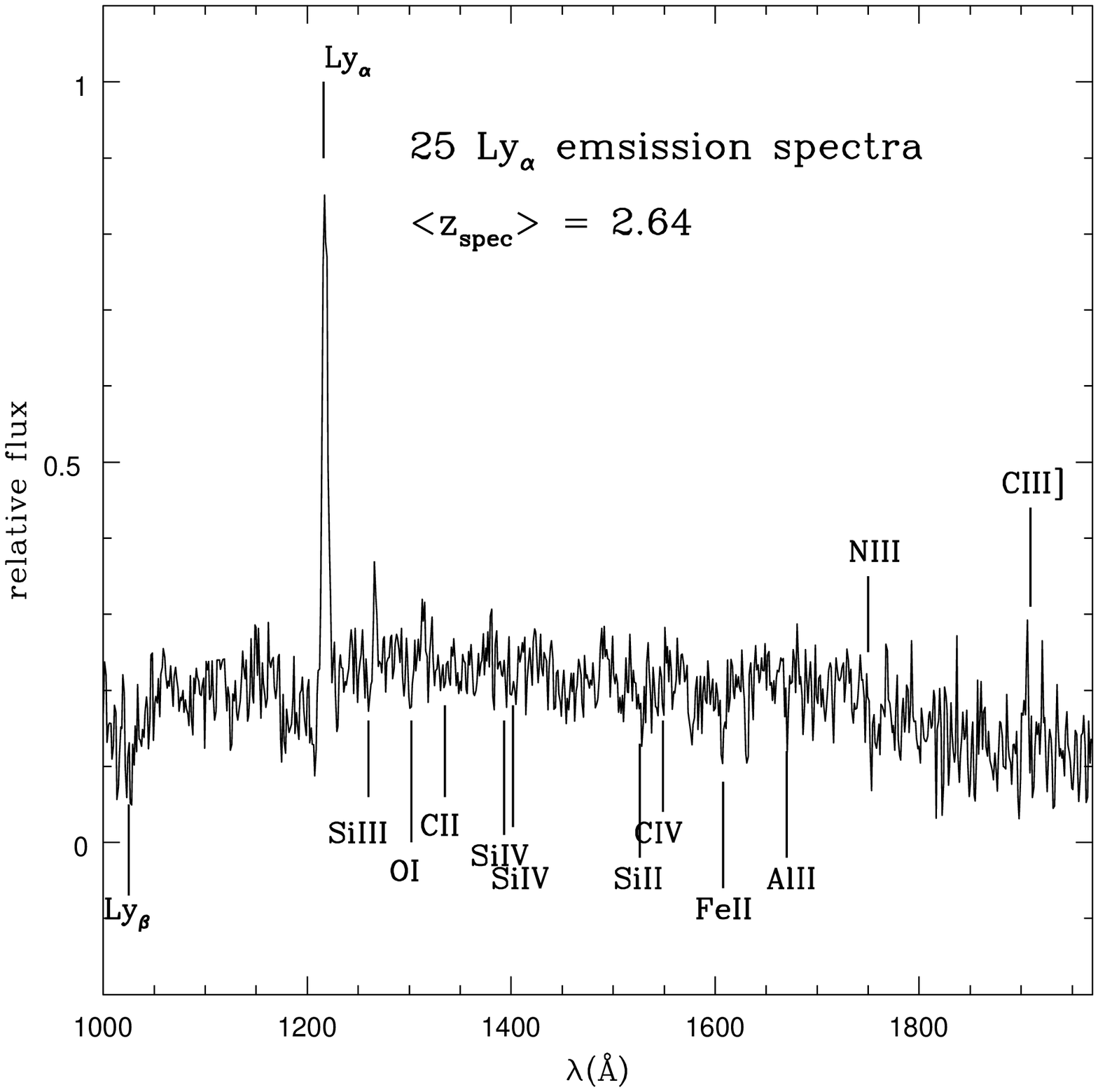}}
\end{minipage}
\begin{minipage}{0.36\textwidth}
\resizebox{\hsize}{!}{\includegraphics{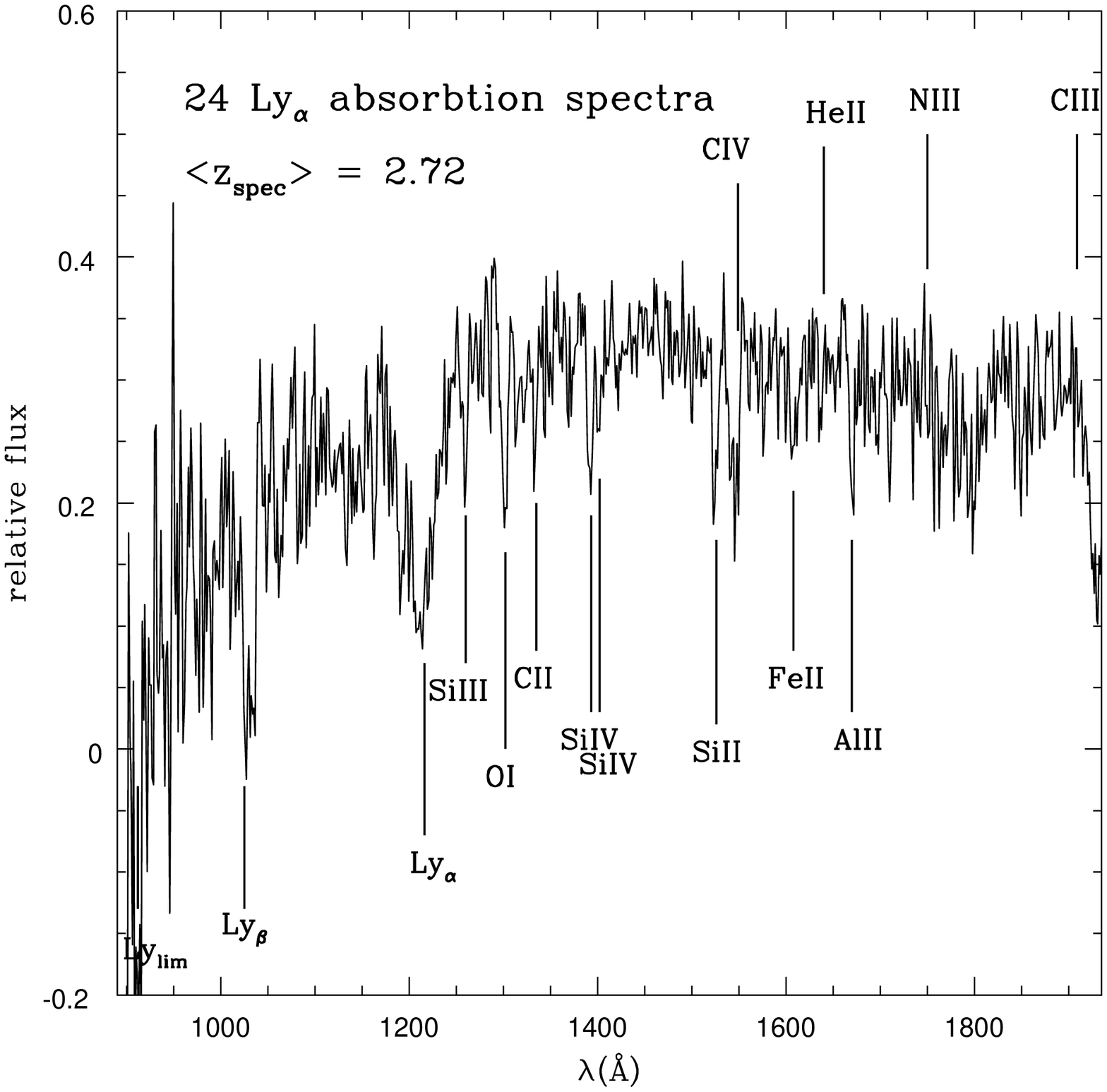}}
\end{minipage}
\begin{minipage}{0.36\textwidth}
\resizebox{\hsize}{!}{\includegraphics{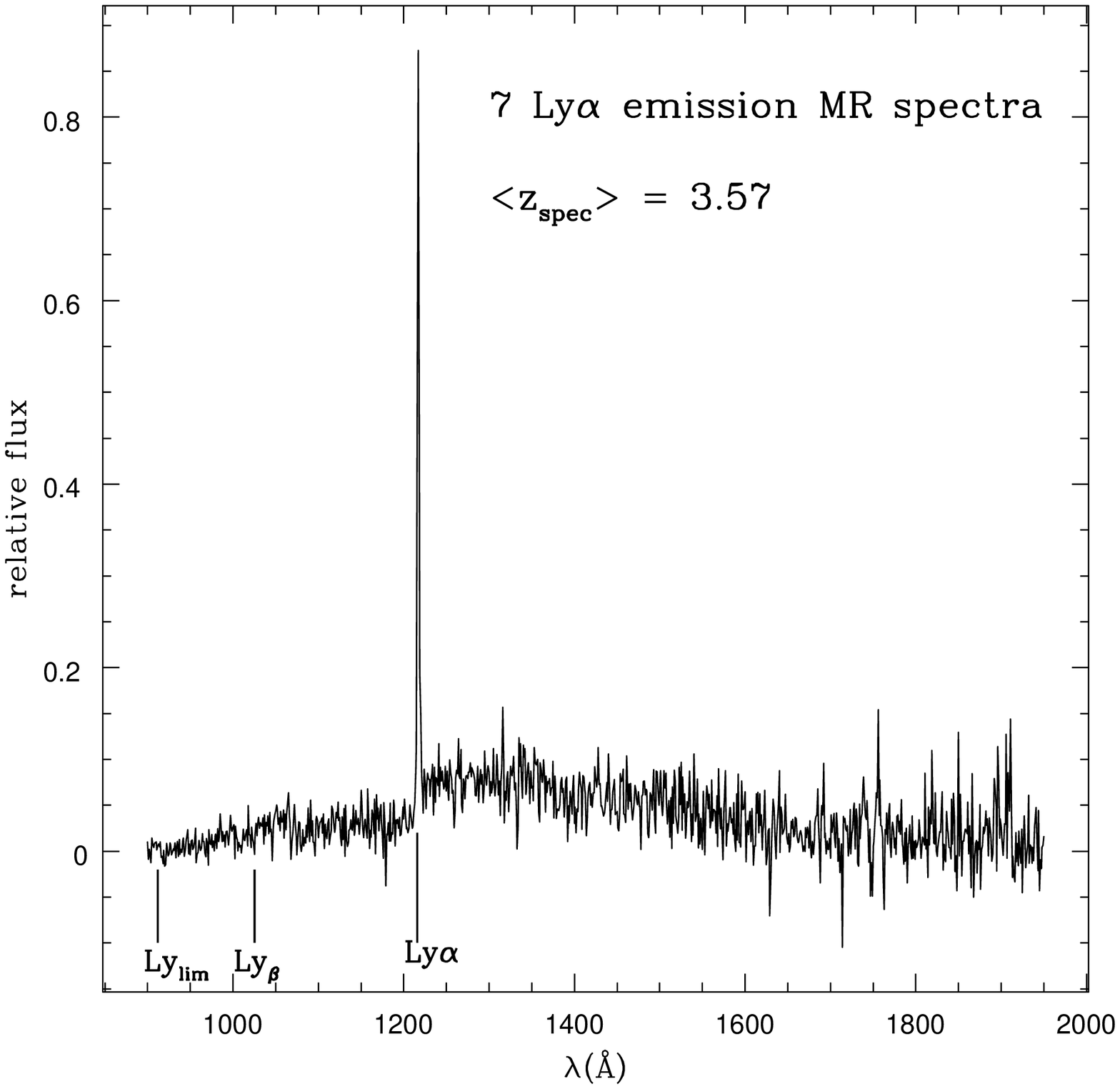}}
\end{minipage}
\begin{minipage}{0.36\textwidth}
\resizebox{\hsize}{!}{\includegraphics{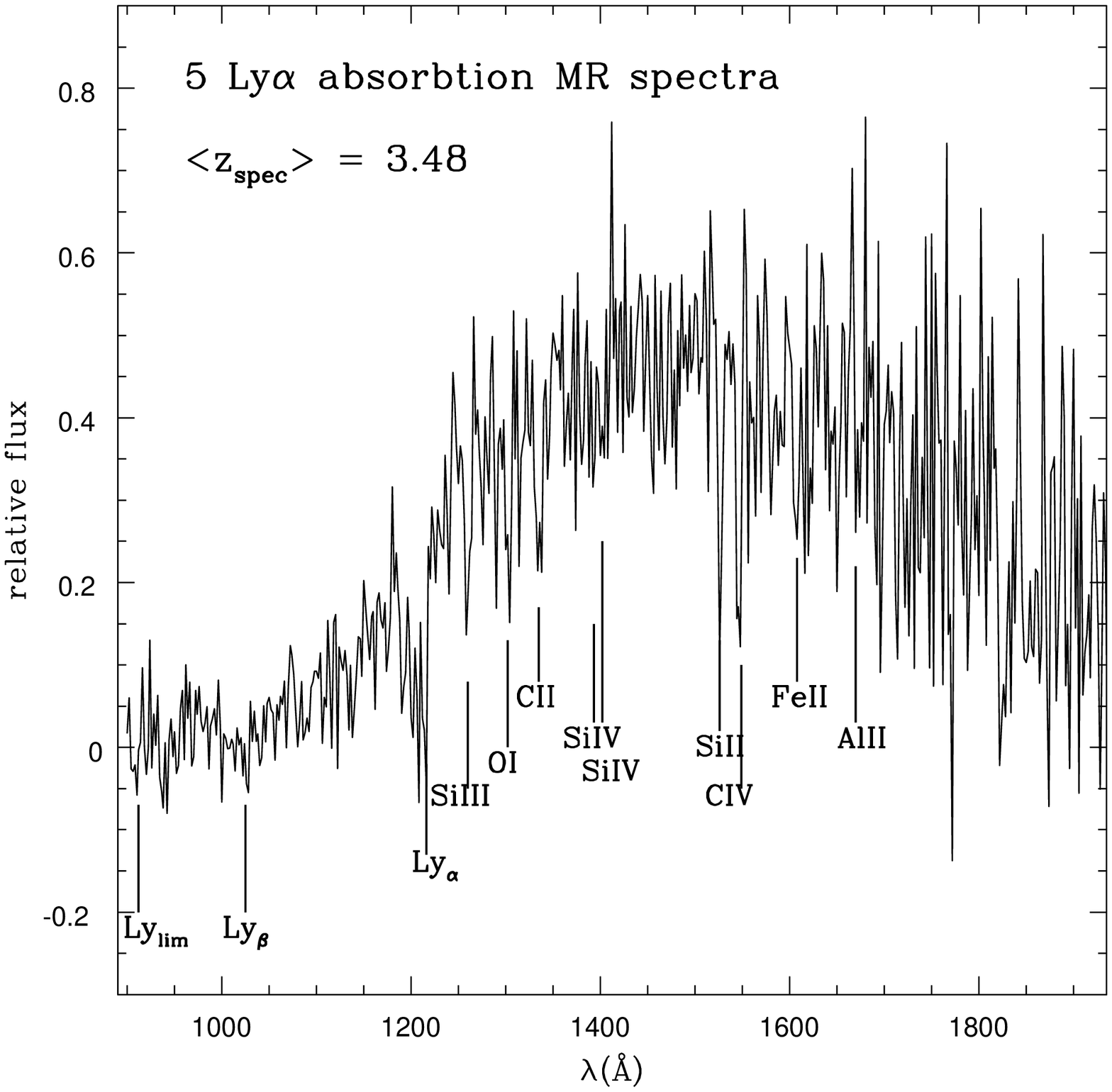}}
\end{minipage}
\begin{minipage}{0.36\textwidth}
\resizebox{\hsize}{!}{\includegraphics{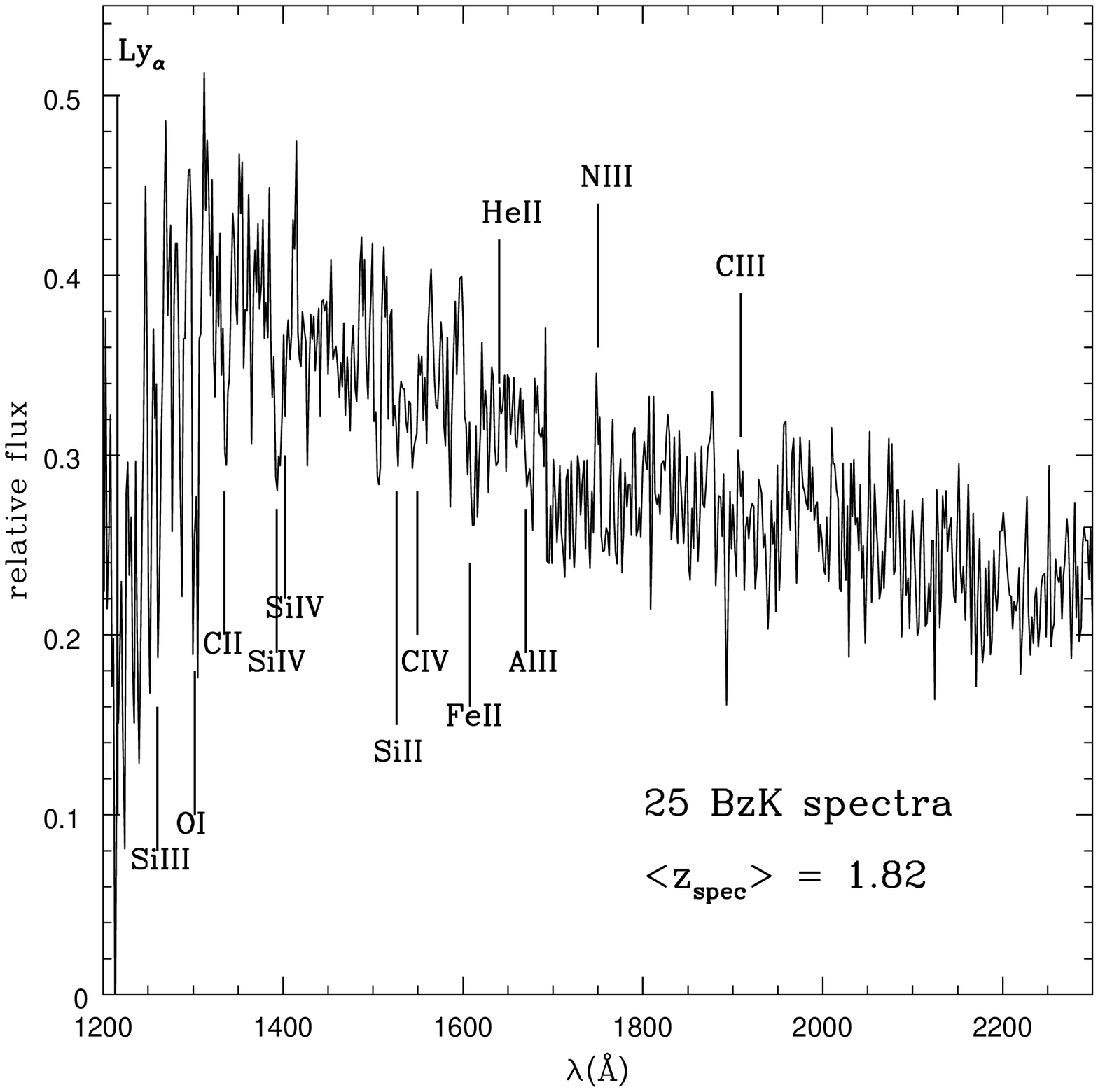}}
\end{minipage}
\end{center}
\caption{Combined 1D spectra of high redshift galaxies in different
redshift bins. The top panels show LBGs with the Ly$\alpha$ in
emission (left panel) and in absorption (right panel) obtained with
the LR-Blue grism at $z \sim 2.7$. The central panels show LBGs with
the Ly$\alpha$ in emission (left panel) and in absorption (right
panel) obtained with the MR grism at $z \sim 3.5$. The bottom panel
shows the combined 1D spectrum of a BzK galaxy obtained with the
LR-Blue grism at $z \sim 1.82$.}
\label{templates}
\end{figure*}

\begin{table}[h]
\caption{Peaks detected in the master catalog redshift distributions,
sorted by increasing redshift. The signal and background distribution
are smoothed with $\sigma_S = 300 km$$s^{-1}$ and $\sigma_B = 15000
km$$s^{-1}$, respectively. Together with the mean redshift of each
peak, the number of sources N within 1000 km$s^{-1}$ from the each
peak and the type of LSS are also shown.}
\begin{center}
\begin{tabular}[h]{cccc}
z & N & S/N &type \\
\hline
      0.1241      &           37  & $>5$ &	groups       \\
      0.2190      &           19  & $>5$ &	sheet like       \\
      0.3393      &           18  & $>5$ &	sheet like       \\
      0.5269      &           26  & $>5$ &	sheet like   \\
      0.6741      &           49  & $>5$ &	filament   \\
      0.7355      &          174  & $>5$ &	cluster   \\
      0.9766      &           31  & $>5$ &	cluster   \\
      1.0310      &           21  & $>5$ &	sheet like   \\
      1.0990      &           45  & $>5$ &	cluster/group   \\
      1.2240      &           48  & $>5$ &	cluster/group   \\
      1.3060      &           13  & $>5$ &	cluster/group   \\
      1.6160      &           13  & $>5$ &	group   \\
      2.3160      &            8  & $>4.5$ &	sheet like   \\
      2.5600      &            7  & $>4.5$ &	sheet like	   \\
\hline
\end{tabular}
\end{center}
\label{tab1}							   
\end{table}

\subsection{Redshift distribution and Large Scale Structure}
Fig. \ref{zdist} shows the fine-grin redshift distribution of the
VIMOS LR-Blue (top panel), the VIMOS MR (the central panel) and the
GOODS master spectroscopic catalog (the bottom panel). The smaller
panels within each main panel show redshift regions of particular
interest. Only the very high quality redshifts have been used for the
analysis (flag A and B VIMOS and FORS2 redshift, flag 1 K20, flag 3
and 4 VVDS redshifts and flag 2 and 3 of Szokoly et al. 2004). To
assess the significance of the observed large scale structures we
follow a procedure suggested by Gilli et al. (2003) and similar to the
one of Cohen et al. (1999). The sources are distributed in
$V=cln(1+z)$ rather then in redshift, since $dV$ corresponds to local
velocity variations granted the Hubble expansion. The observed
distribution is then smoothed with a Gaussian with $\sigma_S=300$
km$s^{-1}$ (see fig. \ref{lss}) to obtain the 'signal'
distribution. Since there is no a priori knowledge of the 'background'
distribution, we heavily smoothed the observed distribution with a
Gaussian with $\sigma_B=15000$ km$s^{-1}$ and considered this as the
background distribution. We then searched for possible redshift peaks
in the signal distribution, computing for each of them the signal to
noise ratio defined as $S/N=(S-B)/B^{1/2}$, where S is the number of
sources in a velocity interval of fixed width $\Delta V=2000$
km$s^{-1}$ around the center of each peak candidate and B is the
number of background sources in the same interval. Adopting the
threshold $S/N \ge 5$ we find 14 peaks.  In order to estimate the
expected fraction of possibly ``spurious'' peaks arising from the
background fluctuations, we have simulated 10$^{5}$ samples of the
same size of the observed distribution and randomly extracted from the
smoothed background distribution and applied our peak detection method
to each simulated sample.  With the adopted threshold, the average
number of spurious peaks due to background fluctuations is 0.09. Of
the simulated samples, 6.6$\%$ show one spurious peak, 0.3$\%$ show
two spurious peaks, and only two simulation (out of $10^{5}$) has
three spurious peaks. None of the simulated samples have four or more
spurious peaks. The 14 peaks detected in the described procedure are
listed in Table \ref{tab1}, with the mean redshift of the peak the
number of object (N) within 1000 km$s^{-1}$ from the peak, the SN
threshold and a short description of the kind of large scale structure
defined by visual inspection of the galaxy spatial distribution. We
briefly compared our findings with previous works:

\begin{itemize}
\item the three clusters at 0.53, 0.67 and 0.73, already seen in the
GOODS-FORS2 and K20 surveys are confirmed by the VIMOS redshifts. The
peak at 0.077 seen in Gilli et al. is not detected in the master
catalog. We do confirm the sheet-like structures observed at 0.219 in
Gilli et al. (2003) and 0.339 (even if they observe a structure at
marginally higher redshift z=0.367).  An additional scale structure is
visible at 0.1241. A cluster-like structure is also visible at 0.9766
as confirmed by an extended X-ray emission observed by Szokoly et
al. (2004). We confirm the detection of the concentrated structures at
z=1,031,1,224,1.616 already seen in K20 by Cimatti et al. (2003), in
the X-ray sample by Gilli et al. (2003) and in the FORS2 sample by
Vanzella et al. (2006). We observe additional significative peaks at
z=1.0990 and 1.3060 as seen by Adami et al. (2005) and Vanzella et
al. (2006).

\item We note that other two peaks have been detected with a
$SN\sim$4.5 at 2.316 and 2.560. The second one has been observed also
by Gilli et al. (2003). In both cases the galaxy within 1000
km$s^{-1}$ from the peak occupy the whole GOODS region in a sheet-like
structure. The probability to detect spurious peaks arising from the
background distribution with a SN equal or greater than SN$\sim$4.5 is
about $10^{-3}$.
\item 124 galaxies are observed in the GOODS master sample in the
redshift range $3 < z < 4$. No over-densities are confirmed in the
considered redshift range.
\item 51 galaxies are observed in the GOODS master sample in the
redshift range $4 < z < 5$ and 46 at $z > 5$. No
over-densities are confirmed in the considered redshift range. The
composite spectra of such absorption and emission high redshift LBGs
are shown in Fig. \ref{templates}.
\end{itemize}

\begin{figure*}
\begin{center}
\begin{minipage}{0.33\textwidth}
 \includegraphics[width=\textwidth]{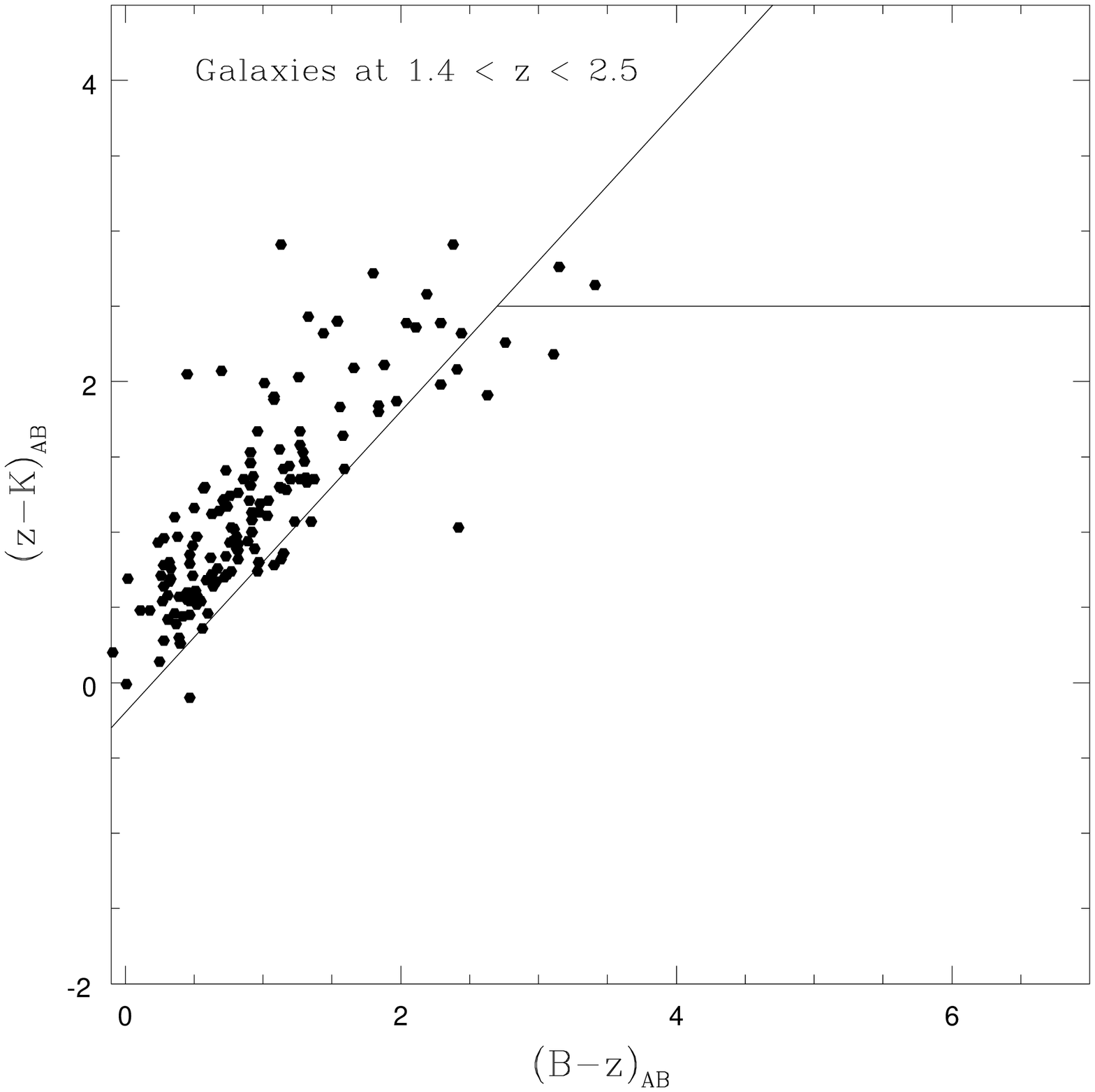}
\end{minipage}
\begin{minipage}{0.33\textwidth}
 \includegraphics[width=\textwidth]{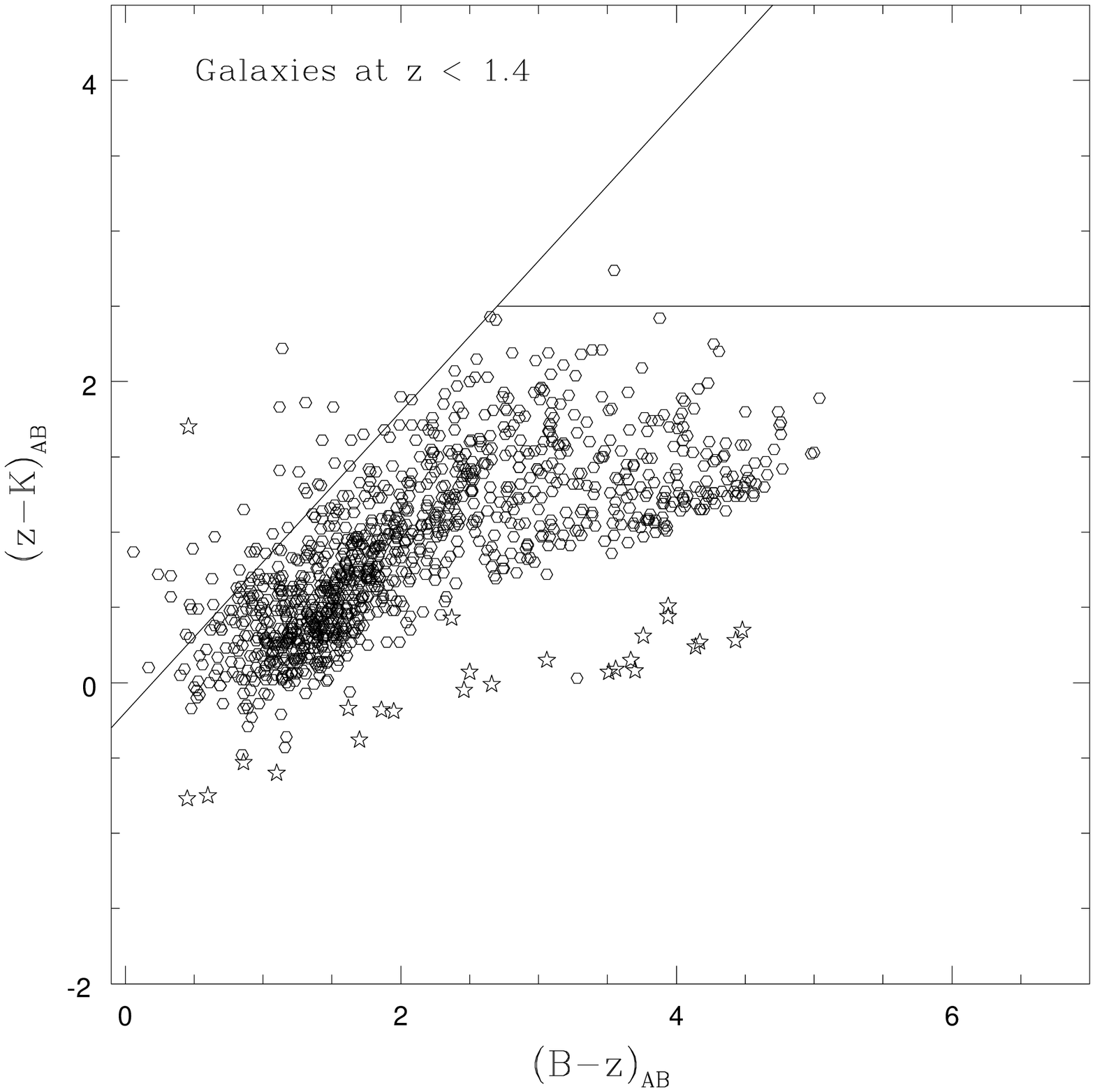}
\end{minipage}
\begin{minipage}{0.33\textwidth}
 \includegraphics[width=\textwidth]{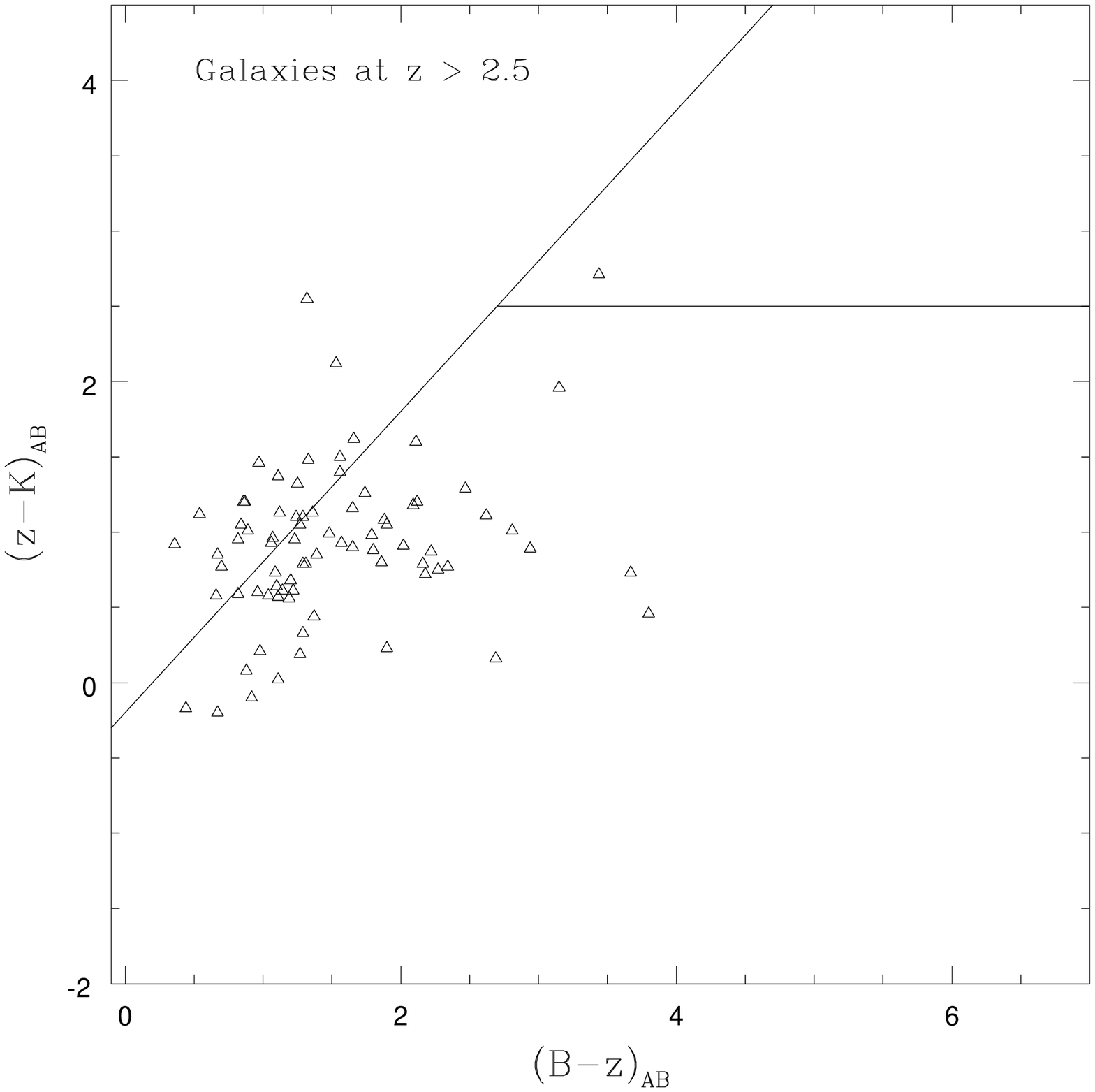}
\end{minipage}
\end{center}
 \caption{BzK diagrams for the GOODS galaxies of the master
 catalog. The left panel shows the galaxies at $ 1.4 < z < 2.5$, the
 central panel shows the low redshift galaxies ($z < 1.4$) and the
 stars, and the right panel shows the galaxies at $z > 2.5$. In the
 central panel, the empty circles refer to the low redshift galaxies and
 the stars refer to the stars. The solid lines show the regions at
 $(z-K)-(B-z) > 0.2$ and $(z-K)-(B-z) < 0.2, B-z > 2.5$ where the star
 forming and passive BzK galaxies, respectively, lie.}
\label{bzk}
\end{figure*}

\begin{figure*}
\begin{center}
\begin{minipage}{0.33\textwidth}
 \includegraphics[width=\textwidth]{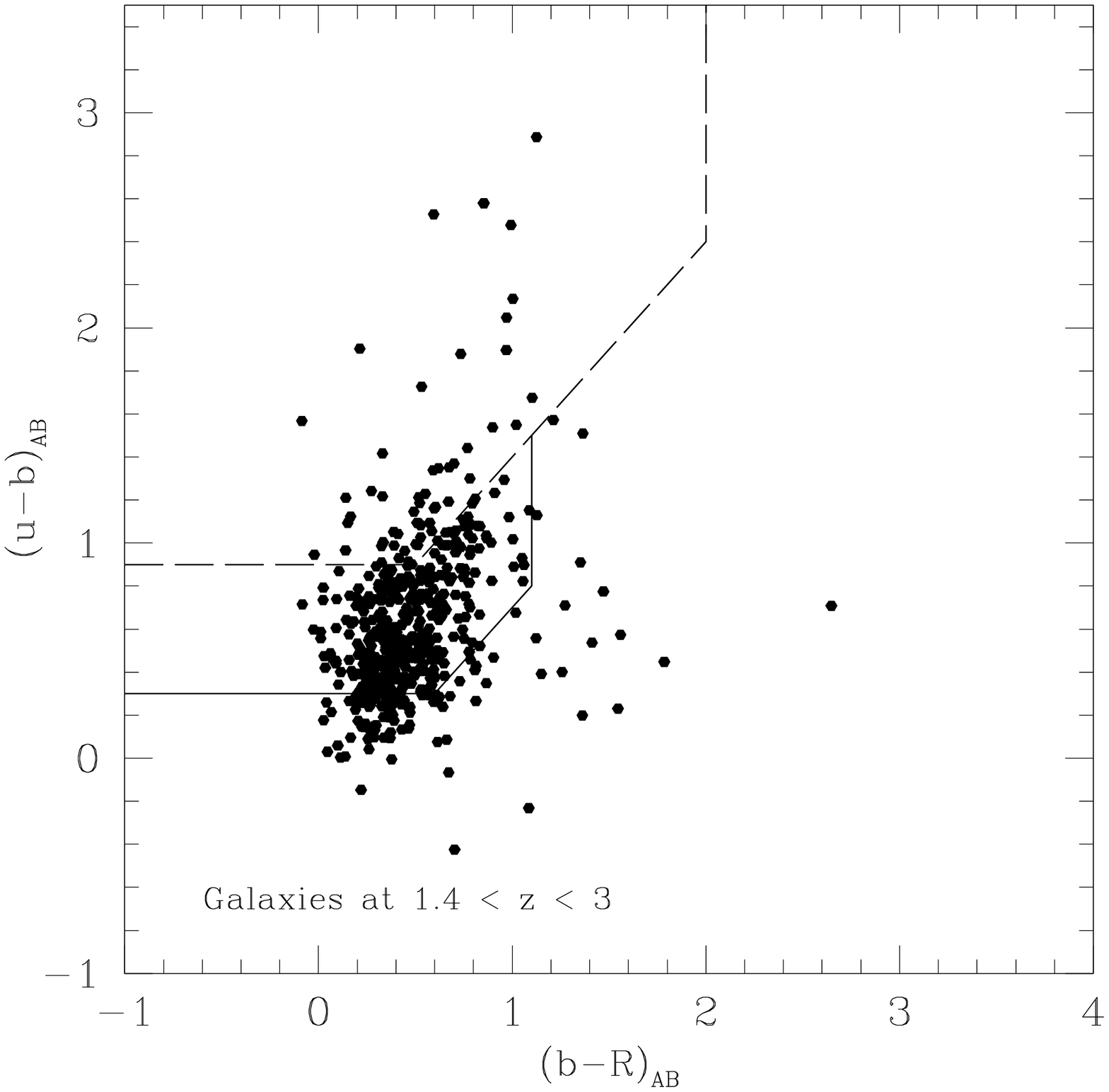}
\end{minipage}
\begin{minipage}{0.33\textwidth}
 \includegraphics[width=\textwidth]{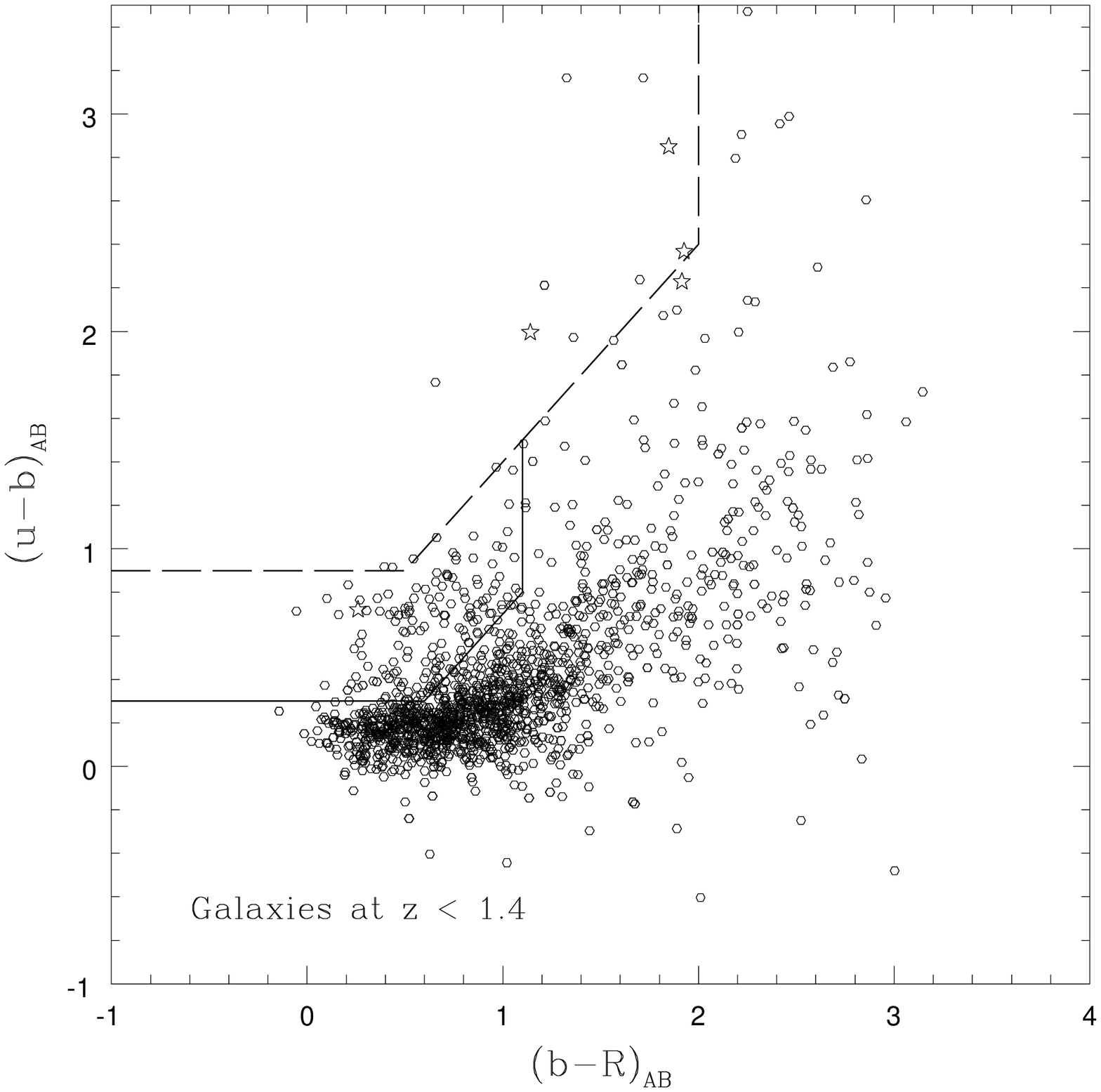}
\end{minipage}
\begin{minipage}{0.33\textwidth}
 \includegraphics[width=\textwidth]{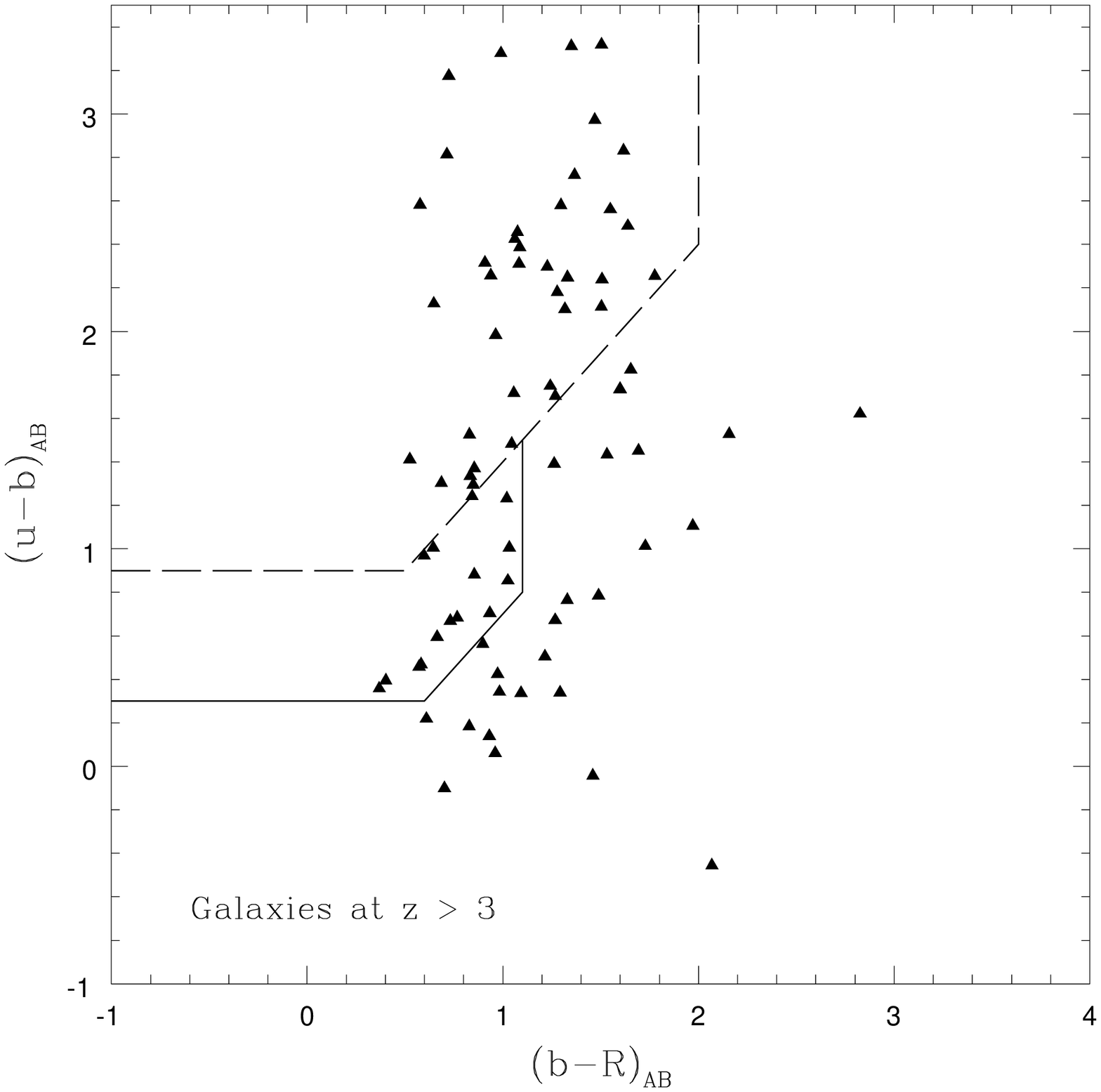}
\end{minipage}
\end{center}
 \caption{$u-b$-$b-R$ color diagrams for the GOODS galaxies of the
 master catalog. The left panel shows the galaxies at $ 1.4 < z < 3$,
 the central panel shows the low redshift galaxies ($z < 1.4$), and
 the right panel shows the galaxies at $z > 3$. In the central panel,
 the empty circles refer to the low redshift galaxies and the stars
 refer to the stars. The region comprised between the solid and the
 dashed lines is the sub-U-dropout locus. The U-dropouts lie above the
 dashed line}
\label{ubr}
\end{figure*}

\begin{figure*}
\begin{center}
\begin{minipage}{0.45\textwidth}
 \includegraphics[width=\textwidth]{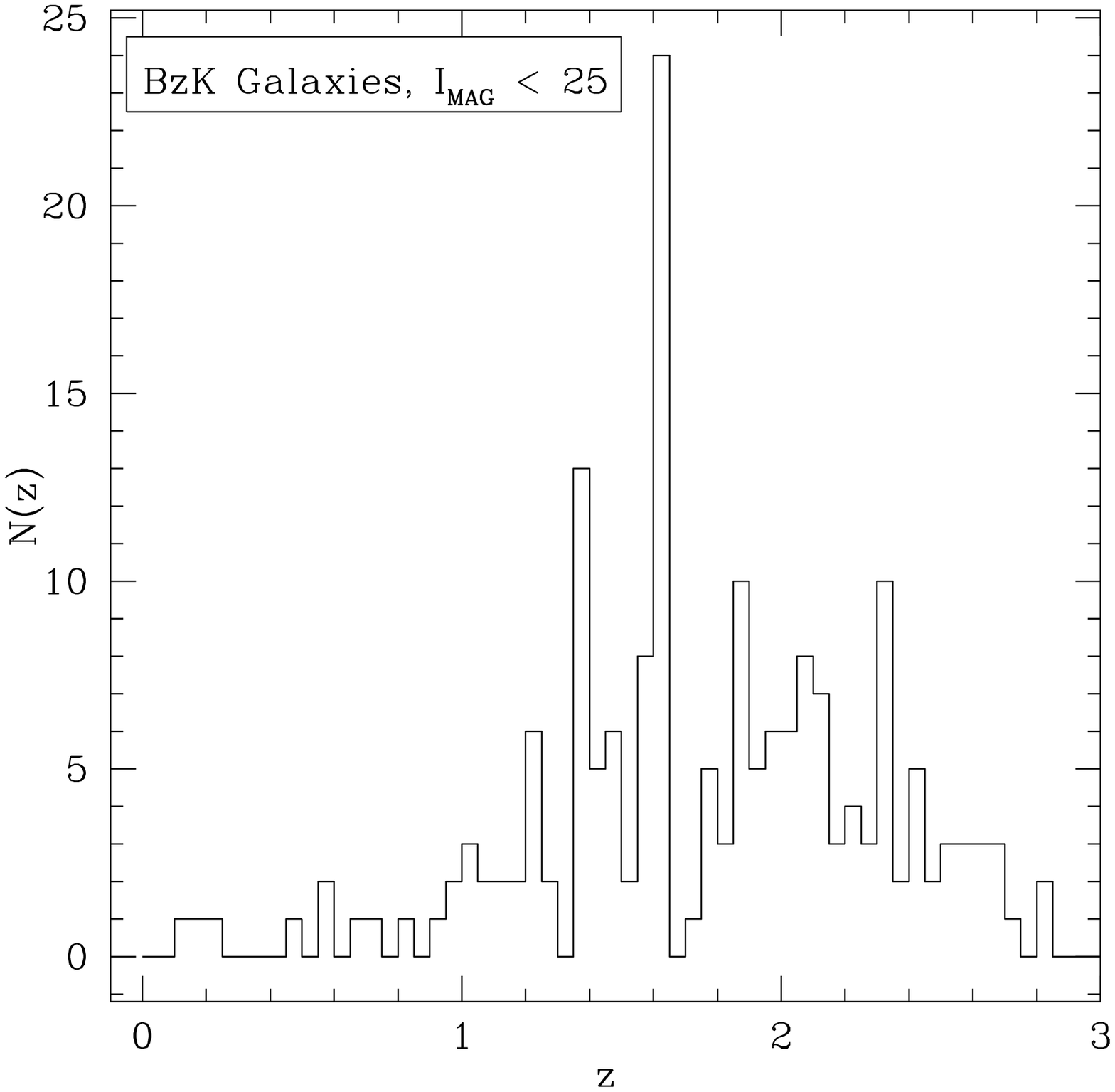}
\end{minipage}
\begin{minipage}{0.45\textwidth}
 \includegraphics[width=\textwidth]{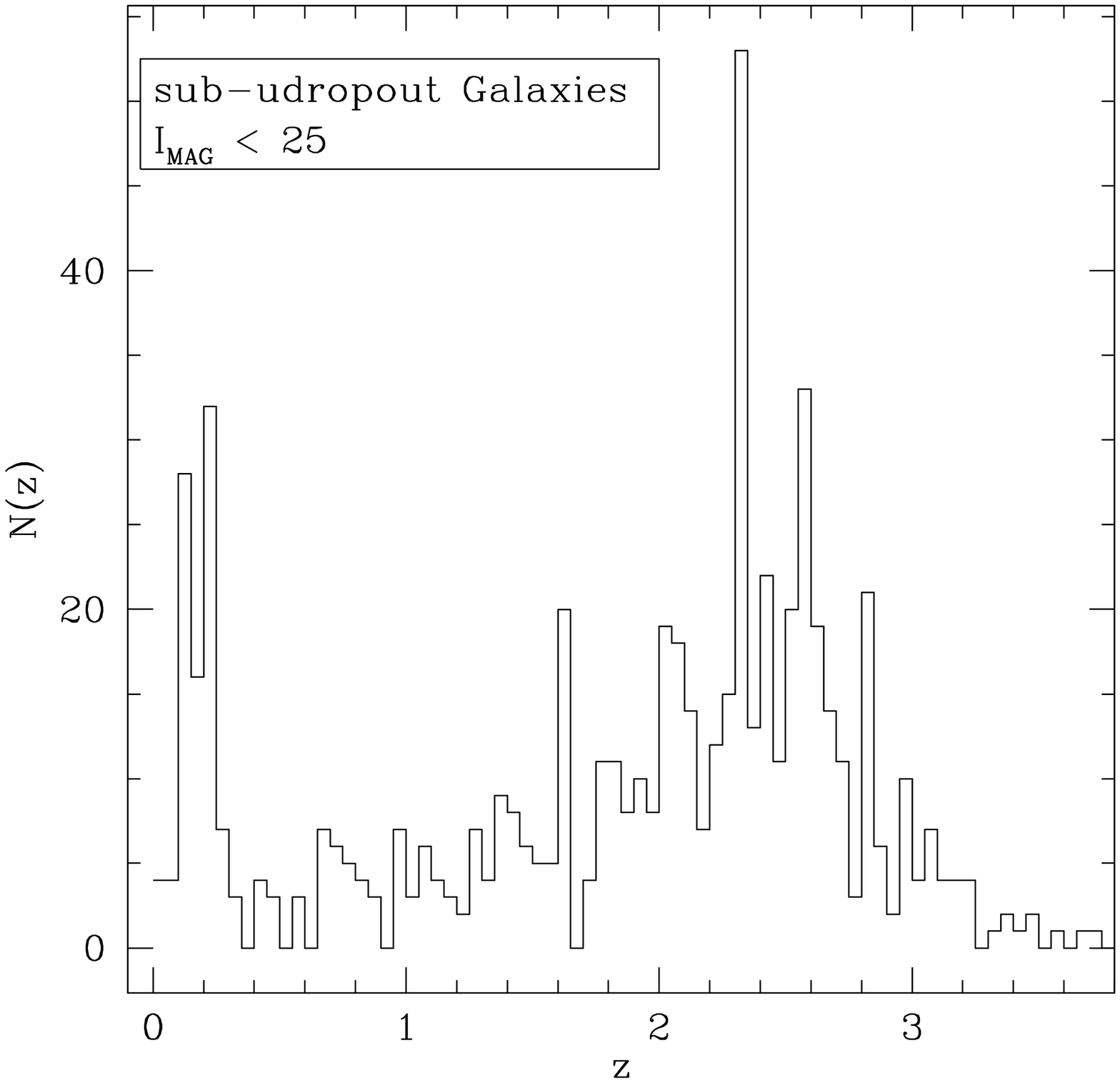}
\end{minipage}
\end{center}
 \caption{Redshift distribution of the BzK selected (left panel) and sub-U-dropout (right panel) galaxies.}
\label{subu}
\end{figure*}

\section{Reliability of the photometric techniques for the selection of galaxies at $1 < z < 3$}

Many photometric techniques have been proposed to select galaxies at
high redshift, namely $z > 1.5$. Many of them (X-ray, BzK,
'sub'-U-dropouts, U, B and V-dropouts criteria) have been used to
target for spectroscopy many of the CDF-S objects in different
spectroscopic surveys. We have combined the results of all these
surveys for creating a master catalog (see previous section), that
reaches a high level of completeness ( $> 50\%$) at $1.5 < z < 3.5$.
The master catalog allows us to check the reliability of different
photometric selection techniques by estimating their contamination due
to low redshift galaxies and their completeness.

Fig. \ref{bzk} shows the BzK diagram of Daddi et al. (2004), which
aims to select galaxies at $z > 1.4$. The galaxy population at $1.4 <
z < 2.5$ (left panel), $z < 1.4$ (central panel) and $z > 2.5$ (right
panel) have been showed in separated diagrams for clarity. The
PSF-matched colors of the BzK catalog used in Daddi et al. (2007a,
2007b) have been used in the diagrams. The solid lines show the
regions at $(z-K)-(B-z) > 0.2$ and $(z-K)-(B-z) < 0.2, B-z > 2.5$
where the star forming and passive BzK galaxies at $z > 1.4$,
respectively, lie. $\sim 86 \%$ of galaxies at $1.4 < z < 2.5$ lie in
the expected BzK region. 14\% of those are not classified as BzK and
lie in the low redshift region. As shown in the central panel, the
whole population (92\%) of the galaxies at $z < 1.4$ populate the
expected BzK region ($(z-K)-(B-z) < 0.2$ and $ B-z < 2.5$) and only
8\% of those are distributed in the other regions of the diagram. The
galaxies at $z > 2.5$ can not be localized in a specific locus in the
diagram. Only 27\% of those higher redshift objects lie at
$(z-K)-(B-z) > 0.2$ and the remaining 73\% are located in the locus of
the low redshift galaxies. This is expected because at $z > 2.5$ the
Lyman forest starts to enter the B band, thus producing a reddening of
the B-z color.  To estimate the contamination due to low redshift
galaxies in the BzK selection, we have distinguished the populations
of BzK star forming ($(z-K)-(B-z) > 0.2$) and passive galaxies
($(z-K)-(B-z) < 0.2, B-z > 2.5$) from the low redshift objects at
$(z-K)-(B-z) < 0.2$ and $B-z < 2.5$ and look for their redshift
distributions. The star forming BzK population has the following
composition: 67\% of the sample lie at $1.4 <z < 2.5$, 10\% at $z >
2.5$ and the contamination of low redshift galaxies is 23\% (see also
the left panel of fig. \ref{subu}). In the BzK passive galaxy region
there are only 2 galaxies. 2 of them are at $1.4 < z < 2.5$, one at $z
> 2.5$ and one at low redshift. We performed this analysis by adopting
different magnitude limit cuts ($I_{MAG}=25$ mag to 23.5 mag), thus
different completeness levels ($\sim 41\%$ to $\sim 55\%$). The
results remain unchanged. This estimate is higher than the $\sim 8\%$
fraction of $z < 1.4$ galaxies found by Daddi et al. (2007). The
difference is largely due to the fact that the latter work excluded
from the analysis hard X-ray sources and blended galaxies.

We have used the same approach to estimate completeness and
contamination of the sub-udrop-out selection criterion.  The purpose
is to select galaxies in the redshift window $1.4 < z < 3$. This
criterion aims to be very similar to the BMBX selection method proposed
by Adelberger et al. (2004). The BMBX selection method is based on
selecting galaxies on the basis of their colors in the $U_n-G$,
$G-R_s$ color color diagram. No $U_nGR_s$ photometry is available for
the CDF-S. So the $U-B$ and $B-R$ colors have been used to define
similar color cuts (see Nonino et al. 2008 in preparation for more
details). The sub-U-dropouts criteria are:
\\
$U-B > 0.3$\\
$U-B > B-R - 0.3$\\
$B-R < 1.1$\\
$R > 23$\\
and not meeting the standard U-drop criterion. The U-dropout criteria are:
\\
$U-B > 0.9$\\
$U-B > B-R + 0.4$\\
$B-R < 2$\\
$R > 23$\\
Fig. \ref{ubr} shows the $u-b$ and $b-R$ color color diagram for the
galaxy populations at $1.4 < z < 3$ (left panel), $z < 1.4$ (central
panel) and $z > 3$ (right panel), respectively. The sub-U-dropout
locus is located below the U-dropouts locus (thus, the name
'sub'-U-dropouts). It is enclosed within the solid and the dashed
lines showed in the diagrams of fig. \ref{ubr}. The U-dropouts lie in
the area enclosed by the dashed line. As showed in the left panel of
the figure, most of the galaxies (80\%) at $ 1.4 < z< 3$ lie in the
locus of the sub-udrop-out region and 8\% in the u-drop locus. Almost
the whole population at $z < 1.4$ (92\%) do not lie in the
sub-U-dropouts and U-dropouts loci, as showed in the central
panel. 58\% of the population at $z > 3$ lie in the U-dropout region
and 16\% in the sub-U-dropout region (left panel). We have estimated
the contamination of the sub-udrop criterion in analogous way to the
BzK method. 72\% of the sub-U-dropouts candidates at $R > 23$ turn out
to be at $1.4 < z < 3$ with a contamination of 24\% of low redshift
objects ($z < 1.4$) and a remaining 4\% of higher redshift objects ($z
> 3$) (see also the right panel of fig. \ref{subu}). The contamination
by low redshift galaxies is similar to that of the BzK selection
method and it is consistent with the results obtained by Adelberger et
al. (2004). We have done the same exercise for the U-dropout (as
defined above) and the B- and V-dropout selection method (as defined
in Giavalisco et al. 2004). In all three cases the LBG technique
provides galaxy samples in the desired redshift range with a $\sim
80\%$ completeness and a $\sim 25\%$ contamination by lower redshift
galaxies.

\section{Conclusions}
In the framework of the Great Observatories Origins Deep Survey a
large sample of galaxies in the Chandra Deep Field South has been
spectroscopically targeted.  A total of 3312 objects with $i_{775}
\mincir 25$ has been observed with the VIMOS spectrograph with the
LR-Blue and MR grism at the ESO VLT providing $2137$ redshift
determinations.  From a variety of diagnostics the measurement of the
redshifts appears to be highly accurate (with a typical $\sigma_z =
0.001$) and reliable (with an estimated rate of catastrophic
misidentification at most few percent).  The spectroscopic coverage of
the CDF-S obtained by combining the VIMOS spectroscopic sample with
the available redshifts in the literature is very high, $\sim 60\%$ up
to redshift $z\sim 3.5$. It is more uncertain above this limit. The
compilation of the redshifts presented in this paper and all redshift
determinations available in the literature have been used to test the
accuracy of the BzK, sub-U-dropout and drop-outs selection
techniques. We showed that any of these methods allows to
create high redshift galaxy samples with a contamination of $\sim
25\%$ of low redshift sources and a completeness level of 80\%. The
same 'master' catalog has been used also to identify several large
scale structures in the GOODS region.

The reduced spectra and the derived redshifts are released to the
community ($\it{http://www.eso.org/science/goods/}$).  They constitute
an essential contribution to reach the scientific goals of GOODS,
providing the time coordinate needed to delineate the evolution of
galaxy properties, morphologies, and star formation and to underhand
the galaxy mass assembly.

\begin{acknowledgements}
  We are grateful to the ESO staff in Paranal and Garching who greatly
  helped in the development of this programme.  We would like to thank
  Martino Romaniello and Carlo Izzo for many stimulating discussions
  and for the help in reducing the VIMOS data. We would like to thank
  also Remco Slijkhuis and Joerg Retzlaff for their work on VIMOS/GOODS
  release.
\end{acknowledgements}

\end{document}